\documentclass[twocolumn,trackchanges]{aastex7}
\usepackage{xspace}
\usepackage[utf8]{inputenc}
\usepackage{upgreek}
\usepackage{hyperref}
\usepackage{amsmath}

\newcommand{\rone}{FRB~20121102A\xspace}
\newcommand{\ronetwin}{FRB~20190520B\xspace}

\newcommand{\rsixseven}{FRB~20201124A\xspace}

\newcommand{\ronefourseven}{FRB~20240114A\xspace}

\newcommand{\rthree}{FRB~20180916B\xspace}
\newcommand{\meightone}{FRB~20200120E\xspace}

\def\torun{Toru\'n\xspace}

\newcommand{\reighteen}{FRB\nobreakspace20190417A\xspace}
\NewPageAfterKeywords

\begin{document}

\title{A milliarcsecond localization associates \reighteen with a compact persistent radio source and an extreme magneto-ionic environment}

\author[orcid=0000-0003-1936-9062,sname='Moroianu']{Alexandra M. Moroianu}
\affiliation{Anton Pannekoek Institute for Astronomy, University of Amsterdam, Science Park 904, 1098 XH, Amsterdam, The Netherlands}
\email[show]{a.m.moroianu@uva.nl}

\author[orcid=0000-0003-3460-506X,sname='Bhandari']{Shivani Bhandari}
\affiliation{SKA Observatory (SKAO), Science Operations Centre, CSIRO ARRC, Kensington WA 6151, Australia}
\email{shivani.bhandari@skao.int}

\author[orcid=0000-0001-7081-0082,sname='Drout']{Maria R. Drout}
\affiliation{David A.~Dunlap Department of Astronomy \& Astrophysics, University of Toronto, 50 St.~George Street, Toronto, ON M5S 3H4, Canada}
\affiliation{Dunlap Institute for Astronomy and Astrophysics, 50 St. George Street, University of Toronto, ON M5S 3H4, Canada}
\email{maria.drout@utoronto.ca}

\author[orcid=0000-0003-2317-1446,sname='Hessels']{Jason W. T. Hessels}
\affiliation{Department of Physics, McGill University, 3600 rue University, Montr\'eal, QC H3A 2T8, Canada}
\affiliation{Trottier Space Institute, McGill University, 3550 rue University, Montr\'eal, QC H3A 2A7, Canada}
\affiliation{Anton Pannekoek Institute for Astronomy, University of Amsterdam, Science Park 904, 1098 XH, Amsterdam, The Netherlands}
\affiliation{ASTRON, Netherlands Institute for Radio Astronomy, Oude Hoogeveensedijk 4, 7991 PD Dwingeloo, The Netherlands}
\email{jason.hessels@mcgill.ca}

\author[orcid=0000-0002-5794-2360,sname='Hewitt']{Dant\'e M. Hewitt}
\affiliation{Anton Pannekoek Institute for Astronomy, University of Amsterdam, Science Park 904, 1098 XH, Amsterdam, The Netherlands}
\email{d.m.hewitt@uva.nl}

\author[orcid=0000-0001-6664-8668,sname='Kirsten']{Franz Kirsten}
\affiliation{Department of Space, Earth and Environment, Chalmers University of Technology, Onsala Space Observatory, 439 92, Onsala, Sweden}
\affiliation{ASTRON, Netherlands Institute for Radio Astronomy, Oude Hoogeveensedijk 4, 7991 PD Dwingeloo, The Netherlands}
\email{franz.kirsten@chalmers.se}

\author[orcid=0000-0001-9814-2354,sname='Marcote']{Benito Marcote}
\affiliation{Joint Institute for VLBI ERIC, Oude Hoogeveensedijk 4, 7991~PD Dwingeloo, The Netherlands}
\affiliation{ASTRON, Netherlands Institute for Radio Astronomy, Oude Hoogeveensedijk 4, 7991 PD Dwingeloo, The Netherlands}
\email{marcote@jive.eu}

\author[orcid=0000-0002-4795-697X,sname='Pleunis']{Ziggy Pleunis}
\affiliation{Anton Pannekoek Institute for Astronomy, University of Amsterdam, Science Park 904, 1098 XH, Amsterdam, The Netherlands}
\affiliation{ASTRON, Netherlands Institute for Radio Astronomy, Oude Hoogeveensedijk 4, 7991 PD Dwingeloo, The Netherlands}
\email{z.pleunis@uva.nl}

\author[orcid=0000-0001-6170-2282,sname='Snelders']{Mark P. Snelders}
\affiliation{ASTRON, Netherlands Institute for Radio Astronomy, Oude Hoogeveensedijk 4, 7991 PD Dwingeloo, The Netherlands}
\affiliation{Anton Pannekoek Institute for Astronomy, University of Amsterdam, Science Park 904, 1098 XH, Amsterdam, The Netherlands}
\email{snelders@astron.nl}

\author[orcid=0000-0002-5519-9550,sname='Sridhar']{Navin Sridhar}
\affiliation{Department of Physics, Stanford University, 382 Via Pueblo Mall, Stanford, CA 94305, USA}
\affiliation{Kavli Institute for Particle Astrophysics \& Cosmology, P.O. Box 2450, Stanford University, Stanford, CA 94305, USA}
\email{navinsridhar@stanford.edu}

\author[orcid=0000-0002-7722-8412,sname='Bach']{Uwe Bach}
\affiliation{Max-Planck-Institut für Radioastronomie, Auf dem Hügel 69, 53121 Bonn, Germany}
\email{ubach@mpifr.de}

\author[orcid=0000-0002-1727-1224,sname='Bempong-Manful']{Emmanuel K. Bempong-Manful}
\affiliation{Jodrell Bank Centre for Astrophysics, Dept.\ of Physics \& Astronomy, University of Manchester, Manchester M13 9PL, UK}
\affiliation{School of Physics, University of Bristol, Tyndall Avenue, Bristol BS8 1TL, UK}
\email{emmanuel.bempong-manful@manchester.ac.uk}

\author[orcid=0000-0003-3655-2280,sname='Bezrukovs']{Vladislavs Bezrukovs}
\affiliation{Engineering Research Institute Ventspils International Radio Astronomy Centre (ERI VIRAC) of Ventspils University of Applied Sciences, Inzenieru street 101, Ventspils, LV-3601, Latvia}
\email{vladislavsb@venta.lv}

\author[orcid=0000-0003-1771-1012,sname='Blaauw']{Richard Blaauw}
\affiliation{ASTRON, Netherlands Institute for Radio Astronomy, Oude Hoogeveensedijk 4, 7991 PD Dwingeloo, The Netherlands}
\email{blaauw@astron.nl}

\author[orcid=0000-0002-0963-0223,sname='Bray']{Justin D. Bray}
\affiliation{Jodrell Bank Centre for Astrophysics, Dept.\ of Physics \& Astronomy, University of Manchester, Manchester M13 9PL, UK}
\email{justin.bray@manchester.ac.uk}

\author[orcid=0000-0002-3341-466X,sname='Buttaccio']{Salvatore Buttaccio}
\affiliation{INAF-Osservatorio Astrofisico di Catania, via Santa Sofia 78, I-95123, Catania, Italy}
\affiliation{INAF-Istituto di Radioastronomia, Via Gobetti 101, 40129, Bologna, Italy}
\email{salvo.buttaccio@inaf.it}

\author[orcid=0000-0002-2878-1502,sname='Chatterjee']{Shami Chatterjee}
\affiliation{Cornell Center for Astrophysics and Planetary Science, Cornell University, Ithaca, NY 14853, USA}
\email{shami@astro.cornell.edu}

\author[orcid=0000-0002-5924-3141,sname='Corongiu']{Alessandro Corongiu}
\affiliation{INAF-Osservatorio Astronomico di Cagliari, via della Scienza 5, I-09047, Selargius (CA), Italy}
\email{alessandro.corongiu@inaf.it}

\author[orcid=0000-0002-9812-2078,sname='Feiler']{Roman Feiler}
\affiliation{Institute of Astronomy, Faculty of Physics, Astronomy and Informatics, Nicolaus Copernicus University, Grudziadzka 5, PL-87-100 Toru\'n, Poland}
\email{rf@astro.umk.pl}

\author[orcid=0000-0002-3382-9558,sname='Gaensler']{B. M. Gaensler}
\affiliation{Department of Astronomy and Astrophysics, University of California, Santa Cruz, 1156 High St., Santa Cruz, CA 95064, USA}
\affiliation{Dunlap Institute for Astronomy and Astrophysics, 50 St. George Street, University of Toronto, ON M5S 3H4, Canada}
\affiliation{David A.~Dunlap Department of Astronomy \& Astrophysics, University of Toronto, 50 St.~George Street, Toronto, ON M5S 3H4, Canada}
\email{gaensler@ucsc.edu}

\author[orcid=0000-0003-4056-4903,sname='Gawroński']{Marcin P. Gawroński}
\affiliation{Institute of Astronomy, Faculty of Physics, Astronomy and Informatics, Nicolaus Copernicus University, Grudziadzka 5, PL-87-100 Toru\'n, Poland}
\email{motylek@astro.umk.pl}

\author[orcid=0000-0002-8657-8852,sname='Giroletti']{Marcello Giroletti}
\affiliation{INAF-Istituto di Radioastronomia, Via Gobetti 101, 40129, Bologna, Italy}
\email{marcello.giroletti@inaf.it}

\author[orcid=0000-0003-2405-2967,sname='Ibik']{Adaeze L. Ibik}
\affiliation{David A.~Dunlap Department of Astronomy \& Astrophysics, University of Toronto, 50 St.~George Street, Toronto, ON M5S 3H4, Canada}
\affiliation{Dunlap Institute for Astronomy and Astrophysics, 50 St. George Street, University of Toronto, ON M5S 3H4, Canada}
\email{ibik@astro.utoronto.ca}

\author[orcid=0000-0002-5307-2919,sname='Karuppusamy']{Ramesh Karuppusamy}
\affiliation{Max-Planck-Institut für Radioastronomie, Auf dem Hügel 69, 53121 Bonn, Germany}
\email{ramesh@mpifr-bonn.mpg.de}

\author[orcid=0000-0002-5857-4264,sname='Lazda']{Mattias Lazda}
\affiliation{David A.~Dunlap Department of Astronomy \& Astrophysics, University of Toronto, 50 St.~George Street, Toronto, ON M5S 3H4, Canada}
\affiliation{Dunlap Institute for Astronomy and Astrophysics, 50 St. George Street, University of Toronto, ON M5S 3H4, Canada}
\email{mattias.lazda@mail.utoronto.ca}

\author[orcid=0000-0002-4209-7408,sname='Leung']{Calvin Leung}
\affiliation{Department of Astronomy, University of California, Berkeley, CA 94720, USA}
\affiliation{Miller Institute for Basic Research, University of California, Berkeley, CA 94720, United States}
\email{calvin_leung@berkeley.edu}

\author[orcid=0000-0002-3669-0715,sname='Lindqvist']{Michael Lindqvist}
\affiliation{Department of Space, Earth and Environment, Chalmers University of Technology, Onsala Space Observatory, 439 92, Onsala, Sweden}
\email{michael.lindqvist@chalmers.se}

\author[orcid=0000-0002-4279-6946,sname='Masui']{Kiyoshi W. Masui}
\affiliation{MIT Kavli Institute for Astrophysics and Space Research, Massachusetts Institute of Technology, 77 Massachusetts Ave, Cambridge, MA 02139, USA}
\affiliation{Department of Physics, Massachusetts Institute of Technology, 77 Massachusetts Ave, Cambridge, MA 02139, USA}
\email{kmasui@mit.edu}

\author[orcid=0000-0002-2551-7554,sname='Michilli']{Daniele Michilli}
\affiliation{Laboratoire d'Astrophysique de Marseille, Aix-Marseille Univ., CNRS, CNES, Marseille, France}
\email{danielemichilli@gmail.com}

\author[orcid=0000-0003-0510-0740,sname='Nimmo']{Kenzie Nimmo}
\affiliation{MIT Kavli Institute for Astrophysics and Space Research, Massachusetts Institute of Technology, 77 Massachusetts Ave, Cambridge, MA 02139, USA}
\email{knimmo@mit.edu}

\author[orcid=0000-0001-9381-8466,sname='Ould-Boukattine']{Omar S. Ould-Boukattine}
\affiliation{ASTRON, Netherlands Institute for Radio Astronomy, Oude Hoogeveensedijk 4, 7991 PD Dwingeloo, The Netherlands}
\affiliation{Anton Pannekoek Institute for Astronomy, University of Amsterdam, Science Park 904, 1098 XH, Amsterdam, The Netherlands}
\email{ouldboukattine@astron.nl}

\author[orcid=0000-0002-8897-1973,sname='Pandhi']{Ayush Pandhi}
\affiliation{David A.~Dunlap Department of Astronomy \& Astrophysics, University of Toronto, 50 St.~George Street, Toronto, ON M5S 3H4, Canada}
\affiliation{Dunlap Institute for Astronomy and Astrophysics, 50 St. George Street, University of Toronto, ON M5S 3H4, Canada}
\email{ayush.pandhi@mail.utoronto.ca}

\author[orcid=0000-0002-5195-335X,sname='Paragi']{Zsolt Paragi}
\affiliation{Joint Institute for VLBI ERIC, Oude Hoogeveensedijk 4, 7991~PD Dwingeloo, The Netherlands}
\email{paragi@jive.eu}

\author[orcid=0000-0002-8912-0732,sname='Pearlman']{Aaron B. Pearlman}
\affiliation{Department of Physics, McGill University, 3600 rue University, Montr\'eal, QC H3A 2T8, Canada}
\affiliation{Trottier Space Institute, McGill University, 3550 rue University, Montr\'eal, QC H3A 2A7, Canada}
\email{aaron.b.pearlman@physics.mcgill.ca}

\author[orcid=0000-0003-2422-6605,sname='Puchalska']{Weronika Puchalska}
\affiliation{Institute of Astronomy, Faculty of Physics, Astronomy and Informatics, Nicolaus Copernicus University, Grudziadzka 5, PL-87-100 Toru\'n, Poland}
\email{wpuchalska@doktorant.umk.pl}

\author[orcid=0000-0002-7374-7119,sname='Scholz']{Paul Scholz}
\affiliation{Department of Physics and Astronomy, York University, 4700 Keele Street, Toronto, ON MJ3 1P3, Canada}
\affiliation{Dunlap Institute for Astronomy and Astrophysics, 50 St. George Street, University of Toronto, ON M5S 3H4, Canada}
\email{pscholz@yorku.ca}

\author[orcid=0000-0002-6823-2073,sname='Shin']{Kaitlyn Shin}
\affiliation{Cahill Center for Astronomy and Astrophysics, MC 249-17 California Institute of Technology, Pasadena CA 91125, USA \\ \hfill \\ \hfill \\ \hfill \\ \hfill \\ \hfill}
\email{kaitshin@caltech.edu}

\author[sname='Sluman']{Jurjen J. Sluman}
\affiliation{ASTRON, Netherlands Institute for Radio Astronomy, Oude Hoogeveensedijk 4, 7991 PD Dwingeloo, The Netherlands}
\email{sluman@astron.nl}

\author[orcid=0000-0002-1530-0474,sname='Trudu']{Matteo Trudu}
\affiliation{INAF-Osservatorio Astronomico di Cagliari, via della Scienza 5, I-09047, Selargius (CA), Italy}
\email{matteo.trudu@inaf.it}

\author[orcid=0000-0001-7361-0246,sname='Williams-Baldwin']{David Williams-Baldwin}
\affiliation{Jodrell Bank Centre for Astrophysics, Dept.\ of Physics \& Astronomy, University of Manchester, Manchester M13 9PL, UK}
\email{david.williams-7@manchester.ac.uk}

\author[orcid=0000-0002-2322-5232,sname='Yang']{Jun Yang}
\affiliation{Department of Space, Earth and Environment, Chalmers University of Technology, Onsala Space Observatory, 439 92, Onsala, Sweden}
\email{jun.yang@chalmers.se}

\begin{abstract}

We report the milliarcsecond localization of a high ($\sim$$1379\rm\,pc\,cm^{-3}$) dispersion measure (DM) repeating fast radio burst, \reighteen. Combining European VLBI Network detections of five repeat bursts, we confirm the FRB's host to be a low-metallicity, star-forming dwarf galaxy at $z = 0.12817$, similar to the hosts of FRBs~20121102A, 20190520B and 20240114A. We also confirm that it is associated with a previously reported persistent radio source (PRS), which is compact on milliarcsecond scales. Visibility-domain model fitting constrains the transverse physical size of the PRS to $<23$\,pc and yields an integrated flux density of $190 \pm 40\rm\,\upmu Jy$ at $1.4$\,GHz. Though we do not find significant evidence for DM evolution, \reighteen exhibits a time-variable rotation measure (RM) ranging between $+3958 \pm 11$ and $+5061 \pm 24\ \rm\,rad\,m^{-2}$ over three years. We find no evidence for intervening galaxy clusters in the FRB's line-of-sight and place a conservative lower limit on the rest-frame host DM contribution of $1228\rm\,pc\,cm^{-3}$ ($90\%$ confidence) --- the largest known for any FRB so far. This system strengthens the emerging picture of a rare subclass of repeating FRBs with large and variable RMs, above-average host DMs, and luminous PRS counterparts in metal-poor dwarf galaxies. Our results suggest that these systems are the result of environmental selection, or a distinct engine for FRB emission.

\end{abstract}

\section{Introduction}
\label{sec:intro}

Fast radio bursts (FRBs) are roughly millisecond-duration, highly luminous ($\gtrsim 10^{42}~\mathrm{erg\,s^{-1}}$) radio transients of extragalactic origin with dispersion measures (DMs) exceeding expectations from Galactic electron density models. FRBs were discovered in 2007 \citep{lorimer2007}, and the development of dedicated high-cadence, wide-field surveys such as CHIME/FRB \citep{chime2018,chime2021cat} have yielded detections of $5,000+$ sources so far (CHIME/FRB Collaboration et al. submitted). Although most FRBs appear as isolated, one-off events, $\lesssim100$ FRBs \citep{chime2023} have been observed to repeat over timescales ranging from seconds to years. The detection of an FRB-like burst from the Galactic magnetar SGR~1935+2154, coincident with an X-ray flare \citep{chime2020magnetar,mereghetti2020, bochenek2020, kirsten2021, li2021, tavani2021, ridnaia2021} showed that at least a fraction of FRBs are produced by magnetars. However, the observed diversity in burst morphologies, repetition rate, polarimetry, and host environments suggests that multiple source classes and/or physical processes may emit FRBs \citep{marcote2020spiral,pleunis2021burstmorph,kirsten2022gc}.

Targeted interferometric observations of active repeating sources have revealed a remarkable diversity in host and local environments. Very long baseline interferometry (VLBI) has enabled parsec-scale localization of repeaters, notably \rthree in the spiral arm of a Milky-Way-like host $\sim$$200\rm\,pc$ offset from a star-forming region \citep{marcote2020spiral, tendulkar2021} and \meightone in a globular cluster in the nearby galaxy M81 \citep{kirsten2022gc}, suggesting a possible origin through binary merger, accretion-induced collapse, or an older compact object progenitor. Polarimetric observations place some repeaters in extreme magneto-ionic environments with rapidly varying rotation measures (RMs) as high as $10^5$\,rad\,m$^{-2}$ \citep{michilli2018} while others are found in much more quiescent environments (e.g., \citealt{mckinven2023rm} and \citealt{ng2025}).

Interestingly, a small subset of these well-localized repeaters are spatially coincident with compact, highly luminous, non-thermal persistent radio sources (PRSs) that are unresolved even at milliarcsecond resolution \citep[see, e.g.,][]{marcote2017r1}. The compact sizes of these sources, coupled with their high radio luminosities \citep{marcote2017r1,niu2022r1twin,bruni2025}, are inconsistent with radio emission expected from star formation. The prototypical source is \rone, the first repeating FRB, which was found to be associated with a compact, luminous ($\sim$$10^{29}\rm\,erg\,s^{-1}\,Hz^{-1}$) PRS embedded in a star-forming, low-metallicity dwarf galaxy \citep{marcote2017r1,chatterjee2017r1,tendulkar2017r1,bassa2017r1}. FRB~20121102A exhibits an extreme, evolving RM ($\sim$$10^4$--$10^5\rm\,rad\,m^{-2}$; source rest frame) consistent with a highly magnetized and dynamic local environment \citep{michilli2018}. A similar case is \ronetwin, which also shows drastic RM variability, even RM sign reversal ($[-3.6,+2.0] \times 10^4\rm~rad\ m^{-2}$; source rest frame), and resides in a comparable host galaxy to \rone. \citep{niu2022r1twin,bhandari2023r1twin,annathomas2023}. These two systems are now considered archetypes of a subclass of FRB-PRS systems that reside in dense, magnetized local environments. 

By contrast, other PRS-associated FRBs such as FRBs~20201124A and 20240114A have shown lower $\rm|RMs|$ ($\sim$$10^2\rm\,rad\,m^{-2}$) and more modest PRS luminosities ($\sim$$ 10^{27}$ -- $10^{28}\rm\,erg\,s^{-1}\,Hz^{-1}$), and remain under active investigation as potential analogues or evolutionary counterparts to FRBs~20121102A and 20190520B\footnote{We note that a compelling candidate PRS has recently been discovered within the CHIME/FRB Outrigger localization region of repeating FRB~20191106C. The repeater exhibits a high RM variability of $-1044.4\pm0.2$ to $-263.3\pm0.2\rm\,rad\,m^{-2}$, while the PRS shares similar properties to those of \rone and \ronetwin, with a specific luminosity of $\sim$$10^{29}\rm\,erg\,s^{-1}\,Hz^{-1}$ and a non-thermal spectral index of $\alpha=-0.60\pm0.05$ \citep{outrigger2025, ng2025}.} \citep{ravi2022, bhusare2024, bruni2024, bruni2025}. An empirical correlation between PRS luminosity and FRB $|\rm RM |$ has been proposed \citep{yang2020}, suggesting that these two properties could serve as evolutionary tracers, with the youngest systems hosting the most luminous PRSs and largest $|\rm RMs|$.

The physical nature of PRSs remains debated. Proposed models include magnetar wind nebulae (MWNe) inflated by relativistic particle outflows \citep{murase2016,metzger2017,margalit2018,margalit2018b}, hypernebulae produced by hyper-Eddington-accreting compact objects \citep{sridhar2022, Sridhar+24}, and accreting wandering massive black holes (MBHs) \citep{annathomas2023,dong2024dwarf}. Models invoking sub-energetic supernovae that leave behind highly magnetized neutron stars (NSs) in dense circumstellar environments are particularly compelling, as they can explain both the persistent emission and high Faraday rotation observed in these systems \citep{metzger2017, margalit2018, piro2018}.

Recently, \citet{ibik2024prs} conducted a targeted search for unresolved PRSs in the roughly arcminute-level localization regions of 37 CHIME/FRB repeaters using archival surveys and targeted Karl G. Jansky Very Large Array (VLA) observations, identifying two promising candidates: PRS~20181030A-S1 and PRS~20190417A-S1. Both exhibit non-thermal spectra and remain unresolved in the VLA images ($2$--$5\arcsec$ resolution), suggesting compact, potentially persistent emission. The latter, PRS~20190417A-S1, is in the field of \reighteen, a repeating FRB with a measured DM of $\sim$$1379\rm\, pc\,cm^{-3}$ (extragalactic component $\sim$$ 1300\rm\,pc\,cm^{-3}$) --- one of the highest known among repeaters \citep{fonseca2020,chime2023}. Its $|\rm RM|$ ($\sim$$4500\rm\,rad\,m^{-2}$ with RM$_{\rm MW}=36\pm13\rm\,rad\,m^{-2}$) is also large, indicating a strongly magnetized local environment consistent with those of the most luminous known FRB-PRS systems \citep{feng2022rm, mckinven2023rm}.

In this Letter, we report the milliarcsecond localization of \reighteen and its PRS, confirming the association proposed by \citet{ibik2024prs} and demonstrating the source's compactness. We present the derived properties of the system, which appears to bridge the gap between the most luminous, high-$|\rm RM|$ FRB-PRS systems and those with more modest properties. We discuss the implications for PRS source models and the environments that give rise to persistent radio emission among repeating FRBs.

\section{Observations \& Data Reduction} 

\subsection{EVN Observations}

We observed \reighteen using the European VLBI Network (EVN) at a central frequency of $1382\rm\,MHz$ as part of the Pinpointing REpeating ChIme Sources with EVN dishes (PRECISE) program\footnote{\url{http://www.ira.inaf.it/precise/Home.html}}. Twenty-five observations were conducted between 2021 October 5 and 2022 August 26 under the project codes EK050 and EK051 (PI: Kirsten). The data were correlated using the FX Software Correlator \citep{kempeima2015} at the Joint Institute for VLBI ERIC (JIVE) with an integration time of 4 seconds and 64 channels per 32-MHz subband. The initial pointing position for the observations was the CHIME/FRB discovery position of \reighteen \citep{fonseca2020}, and subsequent observations were carried out using the CHIME/FRB baseband position \citep{michilli2023baseband}. All initial correlations were performed at the centroid of the baseband localization: $\alpha(\rm J2000) = 19^{\rm h}39^{\rm m}04\arcsec$, $\delta(\rm J2000) = +59\degr19\arcmin55^{s}$, with an uncertainty of $15\arcsec$ in $\alpha$ and $16\arcsec$ in $\delta$ \citep{michilli2023baseband}.

The participating EVN stations were Effelsberg (Ef), Westerbork single dish RT1 (Wb), \torun (Tr), Onsala (O8), Irbene (Ir), Medicina (Mc), Noto (Nt), Urumqi (Ur), and the e-MERLIN stations: Cambridge (Cm), Darnhall (Da), Defford (De), Jodrell Bank Mark II (Jb), Knockin (Kn) and Pickmere (Pi). Scheduling constraints meant that the available antennas varied between epochs. The total observing bandwidth spanned $256\rm\,MHz$, divided into 8 $\times$ 32-MHz subbands. Not all antennas recorded the full bandwidth or were on source for entire sessions, due to differing station capabilities, local sidereal time constraints, and slewing limitations.

Each epoch consisted of alternating 5-minute scans on \reighteen and 1-minute scans on the phase calibrator J1930+5948 (S-band flux density: $0.058\rm\,Jy~beam^{-1}$), located $1.2\degr$ from the target. We also observed J1927+6117 (S-band flux density: $0.646\rm\,Jy~beam^{-1}$) as our primary fringe-finder and bandpass calibrator. A secondary calibrator, J1934+6138 (S-band flux density: $0.211\rm\,Jy~beam^{-1}$), located $2.1\degr$ from the phase calibrator, served as a check source to evaluate the quality of the calibration and assess the final astrometric precision \citep{calibrators2025}. In total, we observed \reighteen for $58.6$\,hr.

Raw baseband voltages recorded at Ef were processed using the PRECISE analysis pipeline\footnote{\url{https://github.com/pharaofranz/frb-baseband}} to search for bursts. The voltages were channelized and transformed into Stokes~I (total intensity) filterbanks \citep{lorimer2011} sampled at $128\,\rm \upmu s$ time resolution with $31.25$\,kHz frequency channels using \texttt{digifil} \citep{vanStraten_2011_PASA}. The intensity data were searched with \texttt{Heimdall}\footnote{\url{https://sourceforge.net/projects/heimdall-astro/}} using a detection threshold of $7\sigma$ and DM range of $1379 \pm 50\rm\,pc\,cm^{-3}$. Burst candidates identified by \texttt{Heimdall} were then classified with \texttt{FETCH} \citep{agarwal2020}, a deep-learning transient classifier; we used the A and H models with a 50\% probability threshold to separate likely astrophysical events from false positives. The final candidates were then inspected by eye to judge their astrophysical origin. The search pipeline is described in detail by \citet{kirsten2021,kirsten2022gc}.

\subsection{Data Calibration \& Imaging}

Post-correlation calibration was carried out in CASA using established VLBI procedures \citep{mcmullin2007casa,casa2022,casa2022fringe}. FITS-IDI files for each observation were retrieved from the EVN data archive\footnote{\url{https://archive.jive.nl/}}, providing two separate data products: (i) continuum files, comprising visibility data for the target and each calibrator; and (ii) burst files, containing $\sim$ms time slices centered on the detected bursts. Calibration metadata were appended using the CASA VLBI extension package, \texttt{casavlbitools}\footnote{\url{https://github.com/jive-vlbi/casa-vlbi/}}, including {\it a priori} amplitude calibration (station gain curves and system temperature measurements), and {\it a priori} flagging tables generated by the EVN AIPS pipeline. The FITS-IDI files were then converted into CASA Measurement Sets (MS) using the \texttt{importfitsidi} task.

Initial data inspection via \texttt{plotms} revealed artifacts from reduced antenna sensitivity in the edge channels, so the first and last four channels per subband were flagged using \texttt{flagdata}. Automated flagging was then applied using the TFCrop algorithm, with standard deviation cuts of $\sigma_{\rm time} = 4$ and $\sigma_{\rm freq} = 3$ to remove time-frequency outliers. Additional manual flagging was performed to remove persistent amplitude or phase artifacts.

Calibration proceeded in three main stages using the \texttt{fringefit} and \texttt{bandpass} tasks: (1) single-band delay calibration to correct for instrumental delays between subbands, (2) bandpass calibration to mitigate frequency-dependent gain variations, and (3) multiband delay calibration, applying global fringe fitting across the full bandwidth. For steps (1) and (2), solutions were derived using the best fringe-finder scan, with a solution determined for each spectral window. Multiband delay calibration used phase-referencing scans to correct for phase variations across the time and frequency domains.

Solutions from all stages were applied incrementally using \texttt{applycal}. The calibrated data for the phase calibrator and check sources were extracted into separate MS files using \texttt{split}. Imaging and iterative self-calibration of the phase calibrator was then performed using \texttt{tclean} and \texttt{gaincal}, producing a refined source model. This improved model allowed for enhanced phase and amplitude calibration across antennas, leading to better calibration of the check source and target.

Dirty images of the bursts and continuum field were generated in CASA using \texttt{tclean} with \texttt{niter=0} and natural weighting to prioritize sensitivity. For the burst, we produced per-baseline delay maps --- signal-to-noise ratio (S/N) as a function of residual delay --- and dirty images to assess the quality of each baseline. Baselines exhibiting a well-defined delay peak or a single, identifiable fringe were classified as high-S/N and retained for subsequent imaging and astrometric analysis. In contrast, noise-dominated baselines with sidelobe ambiguity (i.e., multiple fringe peaks with comparable amplitudes) were excluded to avoid introducing astrometric error and positional bias in the final burst localization.

\section{Analysis \& Results}

\subsection{Burst Properties}

We detected eight bursts from \reighteen over seven of the twenty-five observation epochs of our PRECISE program. We used the FX Software Correlator \citep{kempeima2015} to coherently dedisperse the baseband data of the Ef telescope to $1379$\,pc\,cm$^{-3}$ and create Stokes~I files of each burst. These files have a time and frequency resolution of $0.5\rm\,ms$ and $125\rm\, kHz$, respectively, and are written out as \texttt{sigproc} filterbank files. A plot of the temporal profiles and dynamic spectra of each burst can be found in Figure~\ref{fig:familyplot} (Appendix~\ref{sec:DMappendix}).

The burst properties are summarized in Table~\ref{tab:burstproperties}. The bursts span an order of magnitude in width, from the narrow B2 ($0.60\pm0.1$\,ms) to the two-component B1 ($8.2\pm1.0$\,ms). Burst DMs are determined by maximizing S/N over a grid of trial DMs centered on the fiducial value of $1379$\,pc\,cm$^{-3}$ and fitting a Gaussian to the resulting S/N-DM curve. To ensure reliability, we exclude bursts whose S/N-DM curves are broad or flat due to complex substructure and/or frequency-dependent drift\footnote{More precise DM estimates can be obtained by analyzing the bursts at higher time resolution and optimizing for burst substructure. This analysis is deferred to future work.}. This yields only two bursts (B2 and B5) with robust DM measurements: $1379.2 \pm 2.4$\,pc\,cm$^{-3}$ and $1378.6 \pm 1.5$\,pc\,cm$^{-3}$, respectively. We find a mean DM of $1378.9 \pm 1.4 \rm\,pc\,cm^{-3}$ in the observer frame, consistent with contemporaneous CHIME/FRB DM measurements ($\sim$$1379 \rm\,pc\,cm^{-3}$; see \citealt{curtin2024}). Placing an upper limit of $\Delta\rm DM<2.3 \rm\,pc\,cm^{-3}$ over three years, we find no significant evidence for DM evolution since discovery \citep{fonseca2020}, though an increase of up to $2.3 \rm\,pc\,cm^{-3}$ cannot be excluded.

\begin{table*}
\centering
\caption{Properties of the Bursts Detected from \reighteen in this Campaign.}
\label{tab:burstproperties}
    \begin{tabular}{lccccccccc}
    \hline
         ID & EVN Project & ToA$^{a}$  &S/N$^{b}$&  Width$^{c}$ &  Fluence$^e$& L/I & C/I & RM$_{\text{FDF}}$& RM$_{\text{QU}}$\\
          & Code & (MJD) & & (ms) & (Jy ms) & (\%) & (\%) & (rad\,m$^{-2}$) & (rad\,m$^{-2}$)\\
    \hline
         B1& EK050C & $59619.7519305064$&$10.3$&  $8.2 \pm 1.0^{d}$& $0.44$ & $60 \pm 2$&$6 \pm 2$& $4765\pm27$&-\\
         B2& EK050D & $59654.7669240331$&$11.0$&  $0.6 \pm 0.1$& $0.18$& $98 \pm 9$& $3 \pm 6$ &$5017\pm45$& $5061\pm 24$\\
 B3& EK050F & $59705.5014491605$ &$20.9$& $5.3 \pm 0.3$& $1.13$
& $83 \pm 2$& $2\pm1$ & $3972\pm16$&$3958\pm11$\\
 B4& EK050G & $59789.0391490897$ &$8.3$& $1.6 \pm 0.2$& -&- & - & -&-\\
 B5& EK050G & $59789.0599048523$ &$10.5$& $0.9 \pm 0.1$& $0.15$&$50\pm8$& $-1 \pm 7$ & $4892\pm327$&-\\
 B6& EK051B & $59796.2139888657$ &$7.2$& $3.1 \pm 0.5$& $0.25$&- & -&-&-\\
 B7& EK051A & $59801.1300802646$ &$15.8$& $3.9\pm 0.3$& $0.70$&$45\pm2$& $5\pm2$ &$4907\pm49$&-\\
 B8& EK051D & $59818.1321031052$ &$10.1$& $2.1 \pm 0.3$& $0.28$&$81\pm4$& $5\pm4$ &$4728\pm 37$&$4754\pm13$\\
 \hline
\\
\multicolumn{10}{l}{$^{a}$ Corrected to the Solar System Barycenter (TDB) to infinite frequency assuming a DM of $1379.2\rm\,pc\,cm^{-3}$, a reference frequency}\\
\multicolumn{10}{l}{\quad of $1494\rm\,MHz$ and a dispersion constant of $1/(2.41 \times 10^{-4})\rm\,MHz^2~pc^{-1}~cm^3~s$ at the FRB position quoted in Section~\ref{subsec:R18loc}.}\\
\multicolumn{10}{l}{$^{b}$ This refers to the peak S/N of the time-series.}\\
\multicolumn{10}{l}{$^{c}$ Burst width is measured as the FWHM of a Gaussian fit.}\\
\multicolumn{10}{l}{$^{d}$ B1 is a multi-component burst. The burst width was calculated to encompass both peaks.}\\
\multicolumn{10}{l}{$^e$ We estimate a conservative error of 20$\%$ on these values, which is dominated by the uncertainty on the system equivalent}\\
\multicolumn{10}{l}{\quad flux density of the Ef telescope.}
\end{tabular}
\end{table*}

We perform polarimetric calibration of the burst data from Ef (circular basis) using a test pulsar, PSR~B2255+58, similar to \citealt{kirsten2021}. Thereafter we search a range of RM values between $-10,000$ and $+10,000\rm\,rad\,m^{-2}$ to determine at which RM trial the linearly polarized flux peaks. This is done through RM synthesis, which produces the Faraday dispersion function (FDF) --- a representation of how polarized emission is distributed across Faraday depth \citep{brentjens2005}. We report RM$_{\text{FDF}}$ as the depth corresponding to the strongest peak in this distribution, with values tabulated in Table~\ref{tab:burstproperties}. Three of the bursts (B2, B3 and B8) were sufficiently bright to see the cyclical intensity fluctuations induced by Faraday rotation in Stokes~Q and U across the observing band. For these bursts we also perform a QU-fit \citep{purcell2020} to Stokes Q/L and U/L (where $L = \sqrt{Q^2 + U^2}$ is the total linear polarization) over the spectral extent of the burst (see Appendix~\ref{sec:polarimetry}). We find the RM of \reighteen to be highly variable ($3958 \pm 11$ -- $5061 \pm 24\ \rm\,rad\,m^{-2}$; observer frame), exhibiting fractional variations of $20\%$ and even dropping by $\sim$$1,000\rm\,rad\,m^{-2}$ over a 50-day period (Appendix~{\ref{sec:polarimetry}}, Figure~\ref{fig:rm-evolution}). These results are corroborated by recently published FAST observations, in which 47 bursts were observed from 2021 October 3 to 2022 August 13 with RMs in the range $3946.0$ -- $5225.0\rm\,rad\,m^{-2}$ (observer frame; \citealt{feng2025}).

\subsection{Milliarcsecond Localization of \reighteen}\label{subsec:R18loc}

To mitigate the pointing offset between the CHIME/FRB baseband position \citep{michilli2023baseband} and the true burst position, we adopt an iterative re-correlation scheme similar to the global fringe-fitting implemented in \texttt{CASA}. For each detected burst, we compute the complex lag-spectrum on a high-S/N reference baseline, polarization and spectral window. We then derive the S/N as a function of residual group delay, and fit a one-dimensional Gaussian profile to the S/N-delay curve to obtain the best-fit delay correction. Applying the derived delays to all baselines, we re-correlate the visibilities to a provisional phase center\footnote{${\rm \alpha(\rm J2000)} = 19^{\rm h}39^{\rm m}02\fs16,~ {\rm \delta(\rm J2000)} = +59\degr19\arcmin24\farcs96$.} $\sim$$ 40\arcsec$ from the baseband localization. Imaging the brightest burst (B3) reveals a well-defined ``cross-fringe'' pattern --- the characteristic pattern expected for sparse $(u,v)$-coverage (see, e.g., Figure 4 of \citealt{nimmo2022}) --- with an offset of $\sim$$ 57\arcsec$ from the provisional phase center. Applying this offset, we re-correlate the visibilities to a final, refined phase center\footnote{${\rm \alpha(\rm J2000)} = 19^{\rm h}39^{\rm m}05\fs892, ~{\rm \delta(\rm J2000)} = +59\degr19\arcmin36\farcs99$.}.

Each of the eight bursts from \reighteen is individually imaged, with seven bursts displaying the characteristic cross-fringe pattern. The other burst (B4; see Table~\ref{tab:burstproperties}) is strongly dominated by radio-frequency interference (RFI), with no baselines yielding visible fringes, and is therefore excluded from further analysis. To assess astrometric reliability, we also image the contemporaneous check source data. For five out of seven remaining bursts, the check source positions are consistent across observation epochs, showing modest deviations ($<30\%$ of the beam FWHM) from the cataloged position. These small inter-epoch shifts can be incorporated into the check source positional uncertainty budget, and indicate that the burst localizations for these epochs are robust. In contrast, ionospheric phase variations due to low-elevation ($\lesssim 30\degr$) observations caused significant check source offsets ($>100\%$ of the beam FWHM) for two observational epochs (containing B2 and B3). We omit these bursts from further analysis to avoid introducing systematic errors and ambiguity (i.e., from washing out the peak) in the final burst position.

Our final burst image is constructed by coherently combining the visibilities from the five bursts (B1, B5, B6, B7 and B8) with high S/N and stable calibration, excluding baselines with sidelobe ambiguity. A two-dimensional Gaussian fit to the combined burst visibilities yields a best-fit position for \reighteen within the International Celestial Reference Frame (ICRF):\\\\
$\indent {\rm \alpha_{\rm FRB}\,(J2000)} = 19^{\rm h}39^{\rm m}05\fs8919 \pm4.9\rm\,mas,$\\
$\indent {\rm \delta_{\rm FRB}\,(J2000)} = +59\degr19\arcmin36\farcs828 \pm5.2\rm\,mas.$\\

The quoted positional uncertainties reflect the combined effects of: (i) the formal fitting error derived from the synthesized beam shape and S/N of the detection ($\rm \Delta\alpha=3.60\,mas ,~ \Delta\delta=3.90\,mas$); (ii) uncertainties in the absolute positions of the phase calibrator (J1930+5948; $\pm0.19\rm\,mas$) and check source (J1934+6138; $\pm0.13\rm\,mas$); (iii) the check source positional offset ($\rm \Delta\alpha = 3.04\,mas,~ \Delta\delta=3.09\,mas$); and (iv) an estimate of the frequency-dependent shift in the phase calibrator and check source positions, conservatively $\pm1\rm\,mas$ for each. The derived position is consistent with that of the candidate PRS 20190417A-S1 reported by \citet{ibik2024prs} and confirms that the host galaxy of the candidate PRS is also the host of \reighteen with a PATH probability \citep{aggarwal2021} $P_{\rm PATH}=1.0$.

\subsection{Confirmation of a Compact Persistent Radio Source}\label{subsec:PRS}

We search for the candidate PRS identified by \citet{ibik2024prs} within a $2\arcsec \times 2\arcsec$ continuum field centered on \reighteen. A continuum data set is formed by integrating across the five epochs used for burst imaging, equivalent to $13.6$ hours of on-source time\footnote{In the PRECISE program, baseband data are retained and correlated only for observations in which bursts are detected. For epochs without burst detections, the data are deleted.}. Though we do not remove the burst windows, we find that the combined leakage from the six bursts detected during those epochs (assuming a conservative fluence of $0.3\rm\,Jy~ms$ for the low S/N burst B4) is $0.06\,\upmu\rm Jy$; this is negligible compared to the thermal noise in an EVN continuum image ($\sim$$10\,\upmu\rm Jy$).

The naturally weighted dirty map of the field reveals a single point-like source (S/N $\approx 8$) at the position:\\

\begin{figure}[h!]
\includegraphics[width=1\linewidth]{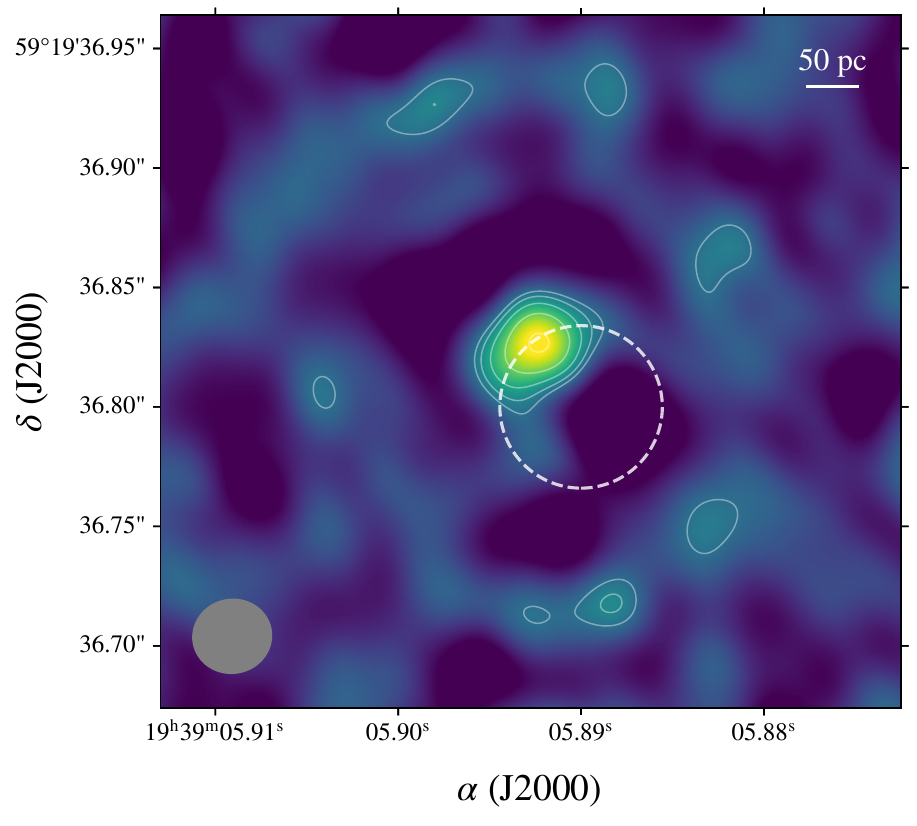}
\caption{EVN dirty map of \reighteen-PRS, as seen in the combined five epochs. A small bar at the top right of the image shows a representative 50-pc transverse extent, for scale. Contour levels start at two times the rms noise level of $12\rm\,\upmu Jy~beam^{-1}$ and increase by factors of $\sqrt{2}$. The dashed circle shows the 1-$\sigma$ VLA positional uncertainty of the PRS 20190417A-S1 \citep{ibik2024prs}. The synthesized beam is represented by the ellipse at the bottom left corner; it has a major and minor axis of $32.7$ and $30.6\rm\,mas$, respectively, and a position angle of $-10.8\degr$.
\label{fig:PRS}}
\end{figure}

\indent ${\rm \alpha_{\rm PRS}\,(J2000)} = 19^{\rm h}39^{\rm m}05\fs8924\pm3.9\rm\,mas$,\\
\indent ${\delta_{\rm PRS}\,(\rm J2000)} = +59\degr19\arcmin36\farcs826\pm4.0\rm\,mas$,\\

referenced to the ICRF. The quoted positional uncertainties are derived using similar principles to those outlined in the previous subsection, though the uncertainty is lower due to the better $(u,v)$-coverage of the continuum observations. Figure~\ref{fig:PRS} shows the continuum image of the compact PRS, which we designate \reighteen-PRS.

Although combining all five epochs nominally increases the integration time, the resulting visibilities are dominated by residual inter-epoch calibration offsets. The epoch corresponding to B4 and B5 exhibited the lowest system temperature and most stable phase solutions, yielding a check source image rms $\sim$$ 2\times$ better than that of the next-best epoch. We therefore use \texttt{Difmap} \citep{difmap1997} to fit a two-dimensional Gaussian to the raw visibilities of the PRS from this single epoch, and place constraints on its angular size and flux density. Our best-fit model (rms $\sim$$ 10\,\upmu\rm Jy\,beam^{-1}$) yields a maximum angular size of $<9.8\rm\,mas$ ($1\sigma$), corresponding to a transverse physical size of $<23\rm\,pc$ at the redshift of the host, $z = 0.12817$. Due to the limited instantaneous bandwidth ($256$\,MHz) and modest S/N of our observations, we are unable to measure a reliable in-band spectral index for \reighteen-PRS. We therefore do not further constrain the value $\alpha = -1.2\pm0.4$ reported by \citet{ibik2024prs}. The total EVN flux density is $190\pm40\,\upmu\rm Jy$ at $1.4\rm\,GHz$, corresponding to a spectral luminosity of $L_{1.4\rm~GHz}=(7.4\pm1.5)\times10^{28}\rm\,erg\,s^{-1}\,Hz^{-1}$. This is consistent with the flux density measurements from \citet{ibik2024prs}; hence, we do not find significant evidence for flux density evolution, though we note our measurements are not tightly constrained.

\begin{figure*}[ht!]
\plotone{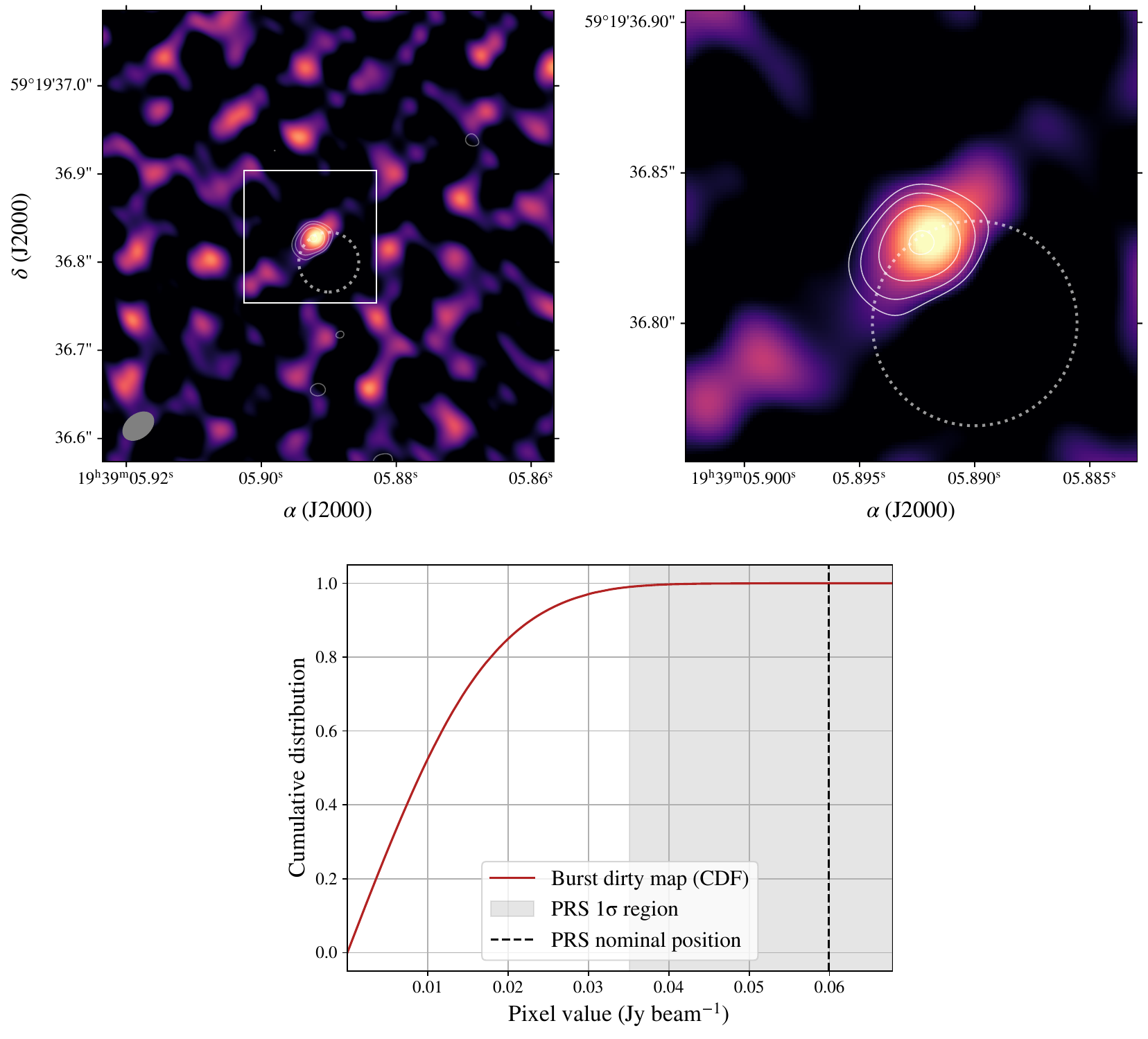}
\caption{\textit{Top left:} $0.5\arcsec \times 0.5\arcsec$ dirty image of five combined \reighteen bursts detected in our EVN-PRECISE observations. The solid contours are taken from Figure~\ref{fig:PRS} and represent the EVN position of \reighteen-PRS, while the dashed circle shows the 1-$\sigma$ VLA positional uncertainty \citep{ibik2024prs}. The synthesized beam is represented by the ellipse at the bottom left corner of the top left panel; it has a major and minor axis of $38.8$ and $26.1\rm\,mas$, respectively, and a position angle of $38.5\degr$. \textit{Top right:} A zoomed-in image on the white square shown in the left panel. \textit{Bottom:} CDF of the pixel values in a zoomed-out $2\arcsec \times 2\arcsec$ field. The black-dotted line shows the pixel value at the position of the EVN PRS centroid, while the gray-shaded region denotes the pixel values within $1\sigma$ of the nominal PRS position.
\label{fig:PRSFRB}}
\end{figure*}

The positions of the combined \reighteen bursts and \reighteen-PRS are consistent within their uncertainties (Figure~\ref{fig:PRSFRB}, top panels). Dirty images of individual bursts obtained with sparsely sampled $(u,v)$-coverage can contain sidelobes with amplitudes close to that of the peak intensity. These sidelobes complicate any assessment of whether another source --- such as the PRS --- falls on the burst source or on a sidelobe maximum. Combining multiple bursts helps remove ambiguity due to the improved $(u,v)$-coverage from Earth rotation. However, to quantify the spatial coincidence between the burst source and PRS, we plot a cumulative distribution function (CDF) of the pixels in a $2\arcsec\times2\arcsec$ dirty image of the burst and investigate how many pixels are brighter than that of the nominal PRS position (Figure~\ref{fig:PRSFRB}, bottom panel). We find that only 0.0031$\%$ of pixels are brighter, ruling out a chance alignment between the two sources ($\approx 4\sigma$). Thus, our EVN localization confirms the existence of a compact radio source within the field of \reighteen and constrains their angular offset to $\Delta\alpha<8.6\rm\,mas,~\Delta\delta<11.0\,mas$, corresponding to a maximum transverse physical offset of 20.3\,pc ($\alpha$) and 26.0\,pc ($\delta$) at the redshift of the host.

\subsection{Host Galaxy Properties}

\citet{ibik2024prs} obtained deep imaging and spectroscopy of an optical source and candidate host coincident with their proposed PRS, demonstrating that it is a star-forming galaxy at $z = 0.12817$. Here we reanalyze these data to constrain additional properties of the galaxy, which we have confirmed is \reighteen's host. The basic reduction of the Gemini imaging and spectroscopy is described by \citet{ibik2024prs}.

\subsubsection{Photometry \& Inferred Galaxy Mass}\label{sec:photom}

Images were obtained in the r-, i-, and z-bands with the Gemini Multi-Object Spectrograph (GMOS; \citealt{GMOS2004}). We supplement this with a g-band image obtained as part of the Dark Energy Camera Legacy Survey (DECaLS; \citealt{Dey2019})\footnote{While the host does not have a magnitude available in the DECaLS catalog, faint emission from the source is visible in the reduced g-band image.}. 
We perform aperture photometry of the host galaxy in all images using the \texttt{iraf} task \texttt{phot}. Because the host is located only $\sim$5.6$\arcsec$ from an m$_r = 15.4$ mag star (which is saturated), the local background is complex. We determine an appropriate sky background by computing the average number of counts in an annulus around the bright star at a similar radius as the PRS host galaxy.
Photometric calibration is then performed based on a set of 6 stars with magnitudes in the PanSTARRS catalog \citep{Flewelling2020}. We find apparent (AB) magnitudes in all four bands of: $m_g = 23.45 \pm 15$ mag, $m_r = 22.42 \pm 0.06$ mag, $m_i = 22.42 \pm 0.07$ mag, and $m_z = 23.32 \pm 0.12$ mag\footnote{We note that this r-band measurement is approximately a magnitude fainter than quoted by \cite{ibik2024prs}. This difference can be attributed to the different means of determining the local background.}.

We then use \texttt{Prospector}\footnote{\url{https://github.com/bd-j/prospector}} \citep{Johnson2021} to fit the revised spectral energy distribution (SED) of the host galaxy.
We use the parametric (exponentially declining) star-formation history model, and find a best-fit stellar mass of $\log_{10}(\rm M_*/M_{\odot})=7.88^{+0.12}_{-0.14}$. Performing synthetic photometry on the best-fit \texttt{Prospector} model, we find an absolute host-rest-frame B-band magnitude of $M_B = -15.5$ Vega mag. Together, these indicate that the host is a faint dwarf galaxy.

\subsubsection{Spectroscopy \& Emission Line Diagnostics}\label{sec:spectrum}

The GMOS spectrum presented by \citet{ibik2024prs} shows a number of strong nebular emission lines. While \cite{ibik2024prs} compute a number of emission line diagnostics, here, we first perform synthetic photometry and scale the spectrum to match the observed i-band magnitude described above (to ensure accurate flux calibration; we chose i-band because it contains the observed H$_{\alpha}$ feature). We then correct for Milky Way (MW) extinction using a \cite{Cardelli1989} extinction curve and $A_{\rm{V}}=0.225$ mag \citep{Schlafly2011}, and shift to host rest frame wavelengths as discussed by \citet{ibik2024prs}. We measure emission line fluxes by fitting Gaussians to the spectrum. Errors are determined via a Monte Carlo approach, creating $1,000$ iterations of the spectrum where each point is sampled from its respective mean and error. Final line fluxes are listed in Appendix~\ref{sec:HostAppendix}, Table~\ref{tab:linefluxes}.

With these revised emission line fluxes, we confirm the results of \citet{ibik2024prs} that the host galaxy falls on the star-forming sequence of the Baldwin, Phillips and Terlevich \citep[BPT;][]{bpt1981} diagram. We measure an H$\alpha$ flux and luminosity of $8.15 \pm 0.02 \times 10^{-16}$\,erg s$^{-1}$~cm$^{-2}$ and $3.57\pm0.01 \times 10^{40}$\,erg\,s$^{-1}$, respectively. These correspond to a star-formation rate of $0.19 \pm 0.01$\,M$_{\odot}$~yr$^{-1}$ using the relation of \cite{Murphy2011}\footnote{This is essentially the same as the value from \cite{ibik2024prs}. Although we find lower apparent magnitudes in Section~\ref{sec:photom} above, the spectrum did not require significant scaling to match the observed photometry}. Coupled with the galaxy stellar mass found above, this implies a specific star-formation rate (sSFR = SFR/M$_*$) of $\sim$$2.3 \times 10^{-8}$\,yr$^{-1}$. Finally, we measure a gas-phase metallicity of $12+\log(\rm O/H)=7.95 \pm 0.01$ using the PP04~N2 diagnostic of \citet[][see Appendix~\ref{sec:HostAppendix} for discussion of other metallicity diagnostics]{Pettini2004}. For a solar value of $12+\log(\rm O/H)=8.69$, this corresponds to slightly less than 20\% solar\footnote{While \citet{ibik2024prs} list a higher (slightly super-solar) metallicity in their Appendix A.9, this was primarily due to a coding error when calculating the metallicity based on the PP04~N2 diagnostic.}.

\section{Discussion}

The milliarcsecond localization of \reighteen marks the confirmation of another FRB-PRS system and offers new insights into the growing population of PRS-associated repeating FRBs.

\subsection{Characteristics of \reighteen}\label{sec:r18discussion}

Taking the observed DM of \reighteen ($\rm 1379.2\,pc\,cm^{-3}$), we decompose it into its constituent parts:

\begin{equation}
\begin{split}
{\rm DM_{obs} = DM_{MW} + \big< DM_{cosmic}(z) \big> } \\\\
+ \big(\sum_i~\frac{\mathrm{DM}_{\mathrm{halo},i}}{1+z_i}\big) + \frac{\mathrm{DM_{host}}}{1+z},
\end{split}
\end{equation}\label{eq:DM}

where $\rm DM_{MW}$ is the Galactic contribution (including the disk and halo), $\big< \mathrm{DM_{cosmic}}\big>$ is the mean intergalactic medium (IGM) contribution, the contributions of intervening halos $\mathrm{DM}_{\mathrm{halo},i}$ are summed over, and $\rm DM_{host}$ represents the contribution from the host galaxy and local environment in the source rest frame. Drawing on the methods outlined by \citet{niu2022r1twin, macquart2020} and \citet{simha2023flimflam}, we derive a distribution for $\rm DM_{host}$ assuming the host galaxy redshift of $z=0.12817$ (see Appendix~\ref{sec:DMappendix} for details).

Our analysis finds a median $\rm DM_{host, rest} = 1275.0\,pc\,cm^{-3}$ and places a $90\%$-confidence lower limit $\rm DM_{host,rest} > 1228.7\,pc\,cm^{-3}$, the largest host contribution yet measured for an FRB\footnote{We note that the measured H$\alpha$ flux could, in principle, be used to constrain the host DM contribution. However, our spectrum provides only a galaxy-averaged flux, introducing large uncertainties in such a conversion. We therefore defer this analysis to future work with spatially resolved H$\alpha$ measurements of \reighteen's local environment.}. As we do not identify any galaxy clusters in the field of \reighteen (Appendix \ref{sec:DMappendix}), such an extreme DM$_{\rm host}$ demands that the FRB source be embedded within, or intersect, a highly dense plasma --- e.g., the ejecta of a young supernova remnant \citep{piro2018}, the regions near an active galactic nucleus (AGN) \citep{zhao2024}, or a compact star-forming \ion{H}{2} region \citep{prayag2024,ocker2024}.

Although we do not find significant DM evolution with $\Delta\rm DM_{obs} < 2.3\rm\,pc\,cm^{-3}$, \reighteen's RM has varied by $\sim$$ 20\%$ over six months (Appendix~{\ref{sec:polarimetry}}, Figure~\ref{fig:rm-evolution}). Consequently, the variable RM must arise from fluctuations in the local magnetic field of the source, as our line-of-sight (LoS) to the FRB is not appreciably varying with time. The corresponding change in the LoS magnetic field integrated through the Faraday medium can be estimated from the RM and DM variance by:

\begin{equation}
    \Delta \big< B_{\parallel} \big> =  1.23 \frac{\Delta \rm RM}{\Delta \rm DM}\rm\,\upmu G.
\end{equation}
\\
Using $\Delta\rm DM_{rest} < 2.6\,pc\,cm^{-3}$ and a net $\Delta \rm RM_{rest} = 1404 \pm 33\rm\,rad\,m^{-2}$ gives us $\Delta \big< B_{\parallel} \big> > 0.65\rm\,mG$.

FAST L-band monitoring of \reighteen between 2021 October and 2022 August, with a fluence completeness threshold of $0.02\rm\,Jy\,ms$, detected 47 bursts with a peak rate of 14\,h$^{-1}$ and mean of 2.26\,h$^{-1}$ \citep{feng2025}. Crucially, \citet{mckinven2023rm} and \citet{feng2025} also report measurable depolarization in individual bursts from \reighteen, suggesting that small-scale fluctuations in $n_eB_{\parallel}$ must be present. Any viable environment for \reighteen must therefore sustain a strong, ordered magnetic field capable of driving large RM variations, while simultaneously allowing for the turbulence or sub-structure that produces the observed depolarization through multi-path propagation (see \citealt{beniamini2022}).

\subsection{Comparison with Known FRBs}

\begin{table*}[ht]
\centering
\caption{Properties of Known FRB-PRS Systems.}
\begin{tabular}{lccccc}
    \hline
Property & \textbf{\reighteen}$^a$ & \rone$^b$ & \ronetwin$^c$ & \rsixseven$^d$ & \ronefourseven$^e$ \\
\hline
$\rm DM_{host,rest}\,(\rm pc\,cm^{-3})$& $> 1228$& $\lesssim 203$& $137$--$707$& $150$--$220$ & $142\pm107$\\
$\rm RM\rm_{rest}\,(rad\,m^{-2})$& $5,038$--$6,441$& $4.4\times10^4$--$1.5\times10^5$& $[-3.6,+2.0]\times10^4$ & $-661\pm42$& $449\pm13$\\
 Peak burst rate$^f$ (h$^{-1}$)& $14$& $122$& -& $542$&$729$\\
$z$ & $0.128$ & $0.193$ & $0.241$ & $0.098$ & $0.130$ \\
$L_{\nu}$ ($\rm erg\,s^{-1}\,Hz^{-1}$) & $\sim$$8\times10^{28}$ & $\sim$$2\times10^{29}$ & $\sim$$3\times10^{29}$ & $\sim$$3\times10^{28}$ & $\sim$$2\times10^{28}$ \\
$\nu$ of above & 1.5\,GHz& 1.4 GHz& 1.7 GHz& 1.6\,GHz& 5 GHz\\
Spectral index, $\alpha$ & $-1.20\pm0.40$ & $-0.15\pm0.08$ & $-0.41\pm0.04$ & $1.00 \pm 0.43$ & $-0.34 \pm0.21$ \\
Physical size\,(pc) & $<23$ & $\leq0.7$ & $<9$ & $<700$ & $<0.4$ \\
PRS-burst offset (pc) & $<26$ & $<40$ & $<80$ & $<188$ & $\sim$$ 28$\\
Host galaxy & Dwarf & Dwarf & Dwarf & Spiral & Dwarf \\
\hline
\end{tabular}\\
\tablerefs{\\ 
$^{a}$ \citet{ibik2024prs}, \citet{feng2025} and this work.\\
$^{b}$ \citet{chatterjee2017r1}, \citet{tendulkar2017r1}, \citet{marcote2017r1}, \citet{li2021_bimodal}, \citet{wang2025} and \citet{bhardwaj2025}.\\
$^{c}$ \citet{niu2022r1twin}, \citet{bhandari2023r1twin}, \citet{annathomas2023} and \citet{lee2023flimflam}.\\
$^{d}$ \citet{hilmarsson2021}, \citet{fong2021}, \citet{piro2021}, \citet{ravi2022}, \citet{nimmo2022}, \citet{zhang2022} and \citet{bruni2024}.\\
$^{e}$ \citet{tian2024}, \citet{bhusare2024}, \citet{zhang2025}, \citet{bruni2025}, \citet{chen2025} and \citet{bhardwaj2025b}.\\
$^{f}$ Peak rates for a fluence completeness threshold of $0.02$\,Jy\,ms.}
\label{tab:frbprsproperties}
\end{table*}

Within the broader FRB population, PRS associations remain rare. While close to 100 repeating FRBs are known, fewer than $5\%$ have been confirmed to be associated with a PRS \citep{bhusare2024,ibik2024prs}. Table~\ref{tab:frbprsproperties} summarizes the properties of known FRB-PRS systems. Hereafter, our discussion will focus on FRBs~20190417A, 20121102A, 20190520B, and 20240114A. Their compact, non-thermal PRSs point to a likely common physical origin. By contrast, the continuum emission co-located with \rsixseven is not compact on milliarcsecond scales \citep{ravi2022,nimmo2022}, and may instead trace dust-obscured star formation in its host, making direct comparison with the four aforementioned FRB-PRS systems difficult \citep{fong2021,dong2024}.

Though the sample size is small, the addition of \reighteen allows us to define some emerging characteristics of these systems:

\begin{itemize}
    \item \textbf{Association with a persistent radio source.} Compact, non-thermal, luminous, and long-lived continuum radio sources co-located with the burst;
    \item \textbf{Extreme, time-variable Faraday rotation.} LoS $|\rm RMs|$ ranging over $10^2 - 10^5\rm\,rad\,m^{-2}$ and varying by tens of percent within months to years;
    \item \textbf{Heightened burst activity.} Peak burst rates exceeding the FRB median ($0.5$--$8\rm\,h^{-1}$ above a fluence completeness threshold of $0.02$\,Jy\,ms; \citealt{feng2025}), often reaching the hyper-active regime ($\sim$ $10^2$\,h$^{-1}$; \citealt{li2021_bimodal,feng2025});
    \item \textbf{Large host-galaxy DM contributions.} DM$_{\rm host}$ ranging over $\sim$ $10^2$ -- $10^3$ $\rm\,pc\,cm^{-3}$;
    \item \textbf{Relation to star formation in metal-poor dwarf galaxies.} Predominance in low-metallicity, star-forming dwarf galaxies. FRBs~20121102A and 20190520B, specifically, reside near star-forming regions within their hosts \citep{bassa2017r1, niu2022r1twin}.
\end{itemize}

\begin{figure*}
    \centering
\includegraphics[trim ={0.7cm 0 1.5cm 0 },clip,width=.48\textwidth]{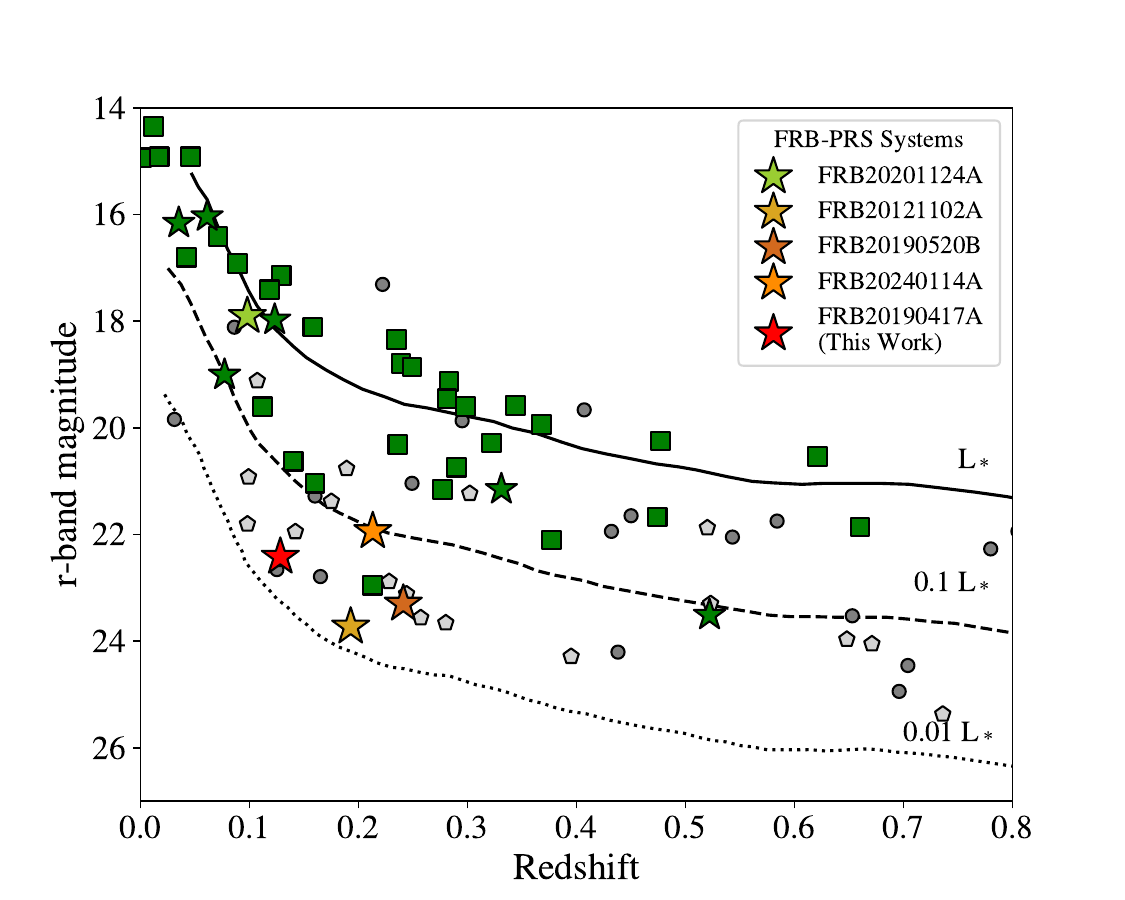}
\includegraphics[trim ={0.7cm 0 1.5cm 0 },clip,width=.48\textwidth]{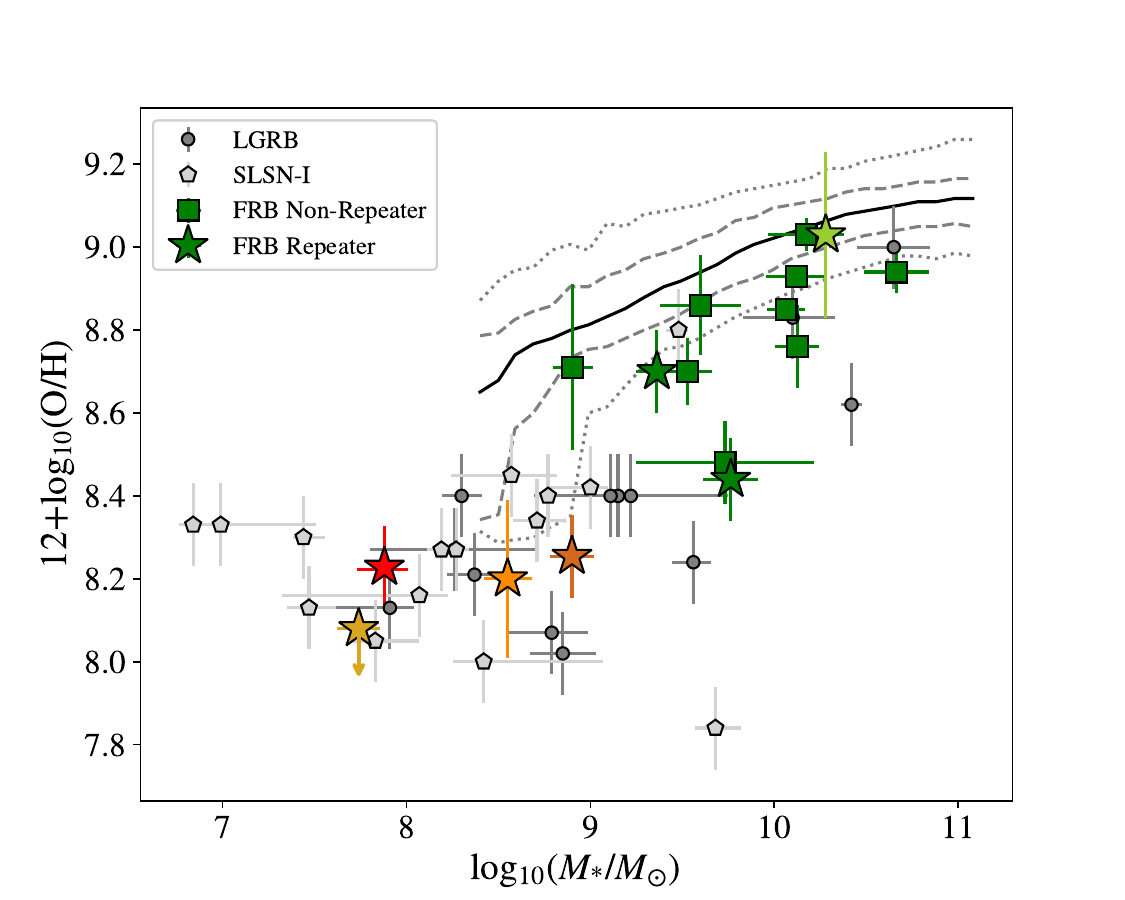}
\caption{The host galaxy of \reighteen in the context of other FRBs and energetic transients. The known FRB-PRS systems (listed in Table~\ref{tab:frbprsproperties}) are shown as colored stars while other FRB hosts are shown as green stars and squares (for repeaters and non-repeaters, respectively). \textit{Left:} Host galaxy apparent r-band magnitude versus redshift. Information for FRB hosts was taken from \cite{Gordon2024} and references therein. For comparison, we also show a set of SLSN-I (pentagons) and LGRB (circles) host galaxies from \cite{Lunnan2014} and \cite{Svensson2010}. The host galaxies of the FRB-PRS systems are notable in that, with the exception of \rsixseven, they are all fainter than $0.1$\,L$_*$ galaxies. \textit{Right:} Galaxy stellar mass versus gas phase metallicity. Black lines represent the median, 68\%, and 95\% contours for SDSS star-forming galaxies from \cite{Tremonti2004}. FRB host galaxy information is taken from \citet{Heintz2020,Bhandari2022} and \citet{Bhardwaj2021}, SLSN-I from \citet{Lunnan2014} and LGRB from \citet{Levesque2010}. The metallicity for the SLSN-I and LGRBs was measured either using an O3N2 diagnostic \citep{Hirschauer2018} or an R23 diagnostic \citep{Kobulnicky2004}. To compare our PP04~N2 measurement from Section~\ref{sec:spectrum}, we convert to R23 using the equations from \citet{Kewley2008}. The host galaxies of the FRB-PRS systems are again notable: they have low masses and metallicities, more comparable to those of SLSN-I and LGRB hosts.}
\label{fig:host-info}
\end{figure*}

These properties differ from those of most FRBs, which show modest $\rm |RMs|\sim$ $5$--$100\rm\,rad\,m^{-2}$ \citep{pandhi2024,ng2025} and host DM contributions $\lesssim 100\rm\,pc\,cm^{-3}$ \citep{khrykin2024,cortes2025}. Typical FRBs also trace the star-forming main sequence of galaxies, showing no strong metallicity preference \citep{yamasaki2025} and a bias toward massive star-forming galaxies \citep{gordon2023,sharma2024}. By contrast, the low-metallicity hosts of FRB-PRS systems (see Figure~\ref{fig:host-info}) mirror the galaxies that give rise to hydrogen-poor super-luminous supernovae (SLSNe-I) and long gamma-ray bursts (LGRBs). This suggests a similar progenitor channel or environmental pathway linking these transients, as first explored by \citet{metzger2017} and \citet{margalit2018b}.

Despite the many parallels between \reighteen and other PRS-associated FRBs, it occupies an intermediate position in terms of its properties --- namely, its $|\rm RM|$ and PRS luminosity. Placing the known systems on the empirical $L_{\nu}-\rm~|RM|$ relation for an expanding magnetized nebula \citep{yang2020} positions \reighteen as a potential bridging source connecting the most magnetized and luminous FRB-PRS systems (\rone, \ronetwin) with weaker ones such as \ronefourseven (see Figure~3 in \citealt{bruni2025}).

\subsection{Physical Interpretation}

It has been proposed that FRB-PRS systems trace the youngest end of the repeating FRB population, consistent with the $L_{\nu}-\rm~|RM|$ relation \citep{yang2020}. In this model, the large and evolving $|\rm RM |$ of \reighteen would reflect a dynamically changing magneto-ionic environment surrounding a young ($10^2$--$10^3$\,yr; \citealt{piro2018}) central engine. However, caution is warranted when employing RM as an evolutionary tracer. The Galactic-center magnetar PSR~J1745$-$2900, for instance, exhibits a large and highly variable RM that is driven by changes in the projected magnetic field/electron column along the LoS rather than by intrinsic evolution \citep{desvignes2018}. Nonetheless, the inferred LoS magnetic field for \reighteen ($\Delta \big< B_{\parallel} \big> > 0.65\rm\,mG$) exceeds typical Galactic values \citep{haverkorn2015} and is instead comparable to extreme, highly magnetized environments like those inferred near Sgr A$^*$ \citep{eatough2013}, the PRS-associated FRBs~20121102A and 20190520B (\citealt{michilli2018,annathomas2023}), and hyper-active repeater FRB~20240619D \citep{ouldboukattine2025}.

While \reighteen's other properties are moderate compared to FRBs~20121102A and 20190520B, its exceptional DM$_{\rm host}$ complicates a straightforward evolutionary scaling. If the host contribution primarily arises from a surrounding nebula, the nebula must be significantly more dense than that of other FRB-PRS systems, pointing to a young nebula in the early stages of expansion. In such scenarios, DM evolution is expected as the nebula expands (though it can be low; \citealt{piro2018}). Within our EVN campaign we find no significant evolution, but we cannot exclude a slight increase ($< 2.3\rm\,pc\,cm^{-3}$) similar to the early behavior of \rone, which initially showed a modest rise in DM \citep{li2021_bimodal,wang2025,snelders2025}. Continued monitoring will be critical for determining whether \reighteen eventually undergoes a DM decline as seen for FRBs~20121102A and 20190520B \citep{hu2023}, or whether its DM remains stable.

\begin{figure*}[h!]
\plotone{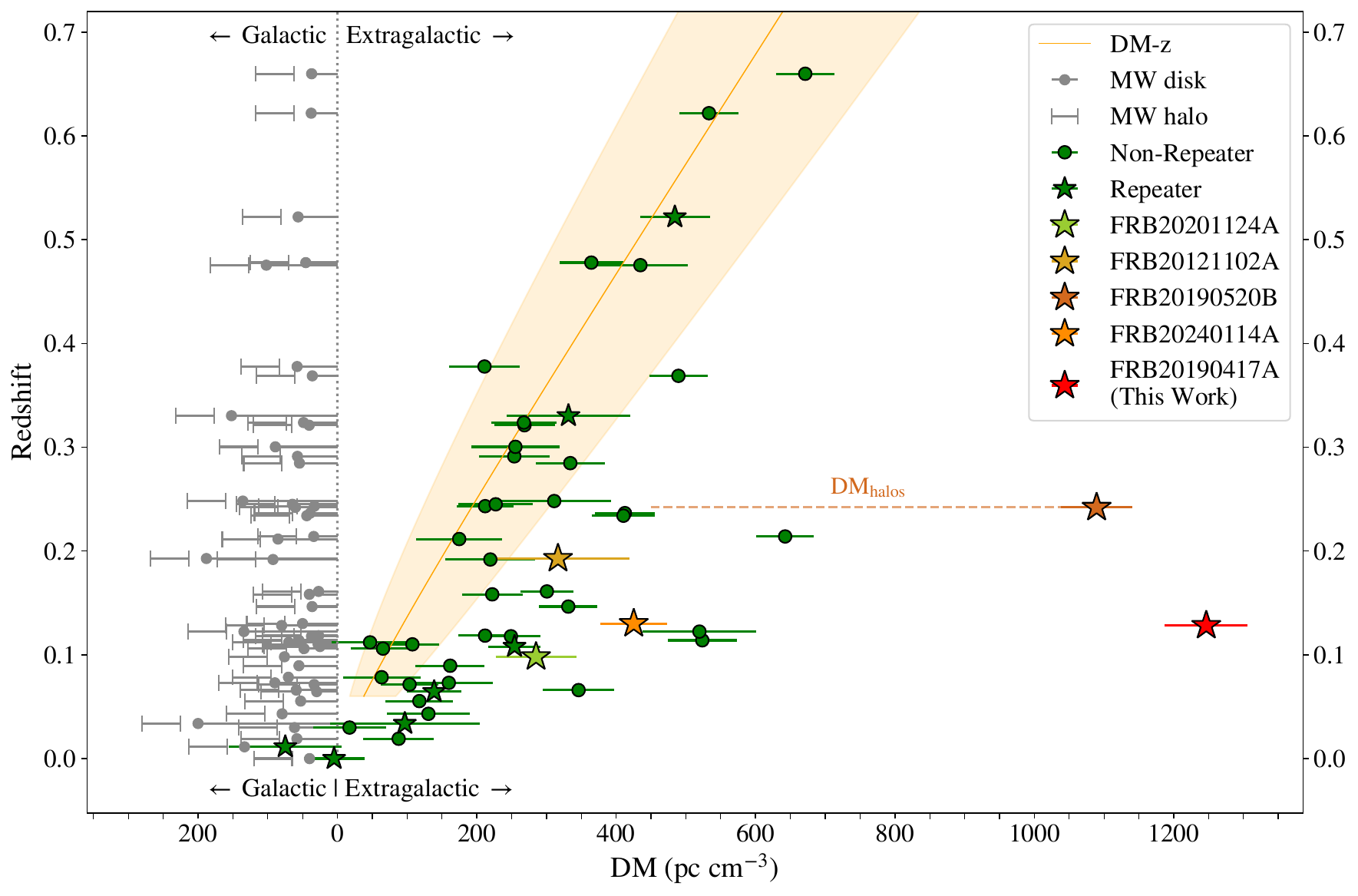}
\caption{Figure~3 from \citet{niu2022r1twin} updated to include $51$ FRBs with robust ($P_{\rm PATH}>0.9$) host associations and redshifts \citep{outrigger2025,law2024}. Galactic disk contributions are estimated from NE2001 with $\pm20\%$ uncertainties, along with an additional halo contribution of $25$--$80$\,pc\,cm$^{-3}$. The red error bars on each extragalactic DM estimate represent a conservative full range uncertainty encompassing the Galactic disk and halo ranges. The expected median DM contribution of the IGM and the inner 1-$\sigma$ confidence interval are given by the orange line and shaded region, respectively. The known PRS-associated FRBs are shown as colored stars while other FRBs are shown as green stars and squares (for repeaters and non-repeaters, respectively). \ronetwin intersects an exceptional LoS \citep{lee2023flimflam}; we include its estimated DM$_{\rm halos}$ contribution for reference.}
\label{fig:DMz}
\end{figure*}

An explanation for \reighteen's extreme DM$_{\rm host}$ is that a large fraction of the excess DM does not originate from its circum-source plasma. Studies of pulsars in the MW and Magellanic Clouds show that dense \ion{H}{2} regions and star-forming knots can add of order $\sim$$10^2\rm\,pc\,cm^{-3}$ to the DM \citep{prayag2024,ocker2024}. One extreme example, PSR~J0248+6021, lies within a dense, giant \ion{H}{2} region, which contributes an excess $\sim$$ 300\rm\,pc\,cm^{-3}$ \citep{theureau2011}. By analogy, \reighteen's exceptional DM$_{\rm host}$ could plausibly result from a sight line intersecting multiple dense \ion{H}{2} regions or star-forming knots, given that these environments are natural birthplaces for young compact objects. More generally, the above-average extragalactic DMs seen in many PRS-associated FRBs (see Table~\ref{tab:frbprsproperties} and Figure~\ref{fig:DMz}) may likewise stem from their location in, or LoS through, such star-forming knots \citep{bassa2017r1,niu2022r1twin}, potentially reflecting high gas fractions in their low-metallicity dwarf hosts.

The metal-poor environments of FRB-PRS systems also have implications for stellar evolution. Reduced line-driven winds in the late evolutionary stages of massive ($\gtrsim8\rm\,M_{\odot}$) OB stars favor the formation of massive, rapidly rotating progenitors that leave behind magnetar or BH remnants \citep{heger2003,monu2006,prestwich2013,aguilera2018,song2023}. For magnetar remnants, rapid rotation enables efficient convective and magneto-rotational dynamos, producing magnetar-strength ($10^{14}$--$10^{15}$\,G) fields \citep{raynaud2020}. The prevalence of FRB-PRS systems in these environments therefore suggests a progenitor channel linking FRBs, SLSNe-I and LGRBs through a common end-of-life outcome: the remnant of a massive, rapidly rotating progenitor embedded in a dense, metal-poor, and highly magnetized circumstellar medium (CSM).

Taken together, these considerations motivate the question: do these systems signify (i) an evolutionary phase, (ii) an environmental selection effect, or (iii) a fundamentally distinct engine for FRB emission?

\begin{itemize}
    \item \textit{Evolutionary phase.} In this picture, every repeater is born inside a luminous, magnetized nebula inflated by a newly formed compact object. As the nebula expands and fades, only the burst source remains, producing the majority population of ``PRS-quiet'' repeaters, i.e., the tail end of the $L_{\nu}-\rm~|RM|$ relation. The rarity of bright PRSs would then reflect the brief youth of the youngest $\sim$$ 5\%$ of repeaters. This scenario is environment-agnostic; luminous PRSs should be found in a variety of FRB host environments.
    \item \textit{Environmental selection.} Alternatively, FRB-PRS systems may form as a consequence of their host environment. Metal-poor, gas-rich dwarf galaxies preferentially produce rapidly rotating, massive stars \citep{schootemeijer2022}. If such a progenitor leaves behind a rapidly spinning compact object (e.g., a millisecond magnetar), the loss of angular momentum over time may cause the compact object to inject its spin-down energy into the surrounding environment, creating persistent and luminous radio emission \citep{metzger2017,margalit2018b}. In this case, FRB-PRS systems should remain confined to the same starburst dwarf environments that host SLSNe-I and LGRBs (Figure~\ref{fig:host-info}).
    \item \textit{Distinct engine.} A third possibility is that PRS-associated FRBs trace a distinct progenitor channel, rather being by-products of youth or host environment. For example, the PRS could be powered by energetic outflows from hyper-Eddington accretion onto a compact object \citep{sridhar2022}, while the FRB is produced along the cleaner jet funnel of the same accreting engine \citep{Sridhar2021}. Alternatively, the PRS could be produced by a wandering MBH, with the FRB originating from a separate compact object embedded in the same environment, or the accretion disk itself \citep{annathomas2023}.
\end{itemize}

At present, the properties of FRB-PRS systems --- particularly, their occurrence in low-metallicity, star-forming dwarf galaxies --- appear to favor either an environmental selection effect, or a distinct engine origin for PRSs.

\subsection{The Nature of PRSs}

The discovery of PRSs has given rise to a number of theoretical models to explain their nature and association with repeating FRBs. The most prominent theories invoke MWNe, hypernebulae powered by hyper-Eddington accreting compact object binaries, and wandering MBHs.

\subsubsection{Young Magnetar Wind Nebula}

In this model, the FRB source is a young magnetar embedded within the expanding debris of its birth supernova. Energy injected via spin-down magneto-hydrodynamic winds or episodic magnetar flares inflate a magnetized nebula inside the supernova remnant shell; synchrotron radiation from this nebula produces the observed luminous PRS \citep{metzger2017,margalit2018,margalit2018b}. The surrounding magneto-ionic ejecta/CSM can yield the large, variable RMs and high DM$_{\rm host}$ typical of PRS-associated FRBs; nebular expansion and shocks also account for the evolving DMs of FRBs~20121102A and 20190520B \citep{piro2018,li2021_bimodal,hu2023,wang2025,snelders2025}. VLBI PRS size constraints ($\sim$ $0.4$--$23.1$\,pc; Table~\ref{tab:frbprsproperties}) are consistent with nebular expectations --- for reference, the Crab Nebula, a $\sim$$10^3\,$yr old pulsar wind nebula (PWN), has a diameter of $3.4$\,pc --- and the rarity of PRS-associated FRBs is compatible with the young ages ($t_{\mathrm{age}} \lesssim 10^3$\,yr) implied by a MWN origin \citep{piro2018,margalit2018,bhattacharya2024,rahaman2025}.

Another strength of the MWN model is that all confirmed FRB-PRS systems are consistent with the empirical $L_{\nu} - \rm |RM|$ relation predicted for evolving nebulae \citep{yang2020}. Modelling efforts point to two distinct MWN regimes:

\begin{itemize}
    \item \textit{Rotation powered.} \citet{bhattacharya2024} show that rotation-powered MWNe with $t_{\mathrm{age}} \approx 20$\,yr and initial spin periods of $P_i \approx 1-3$\,ms can reproduce the properties of FRBs~20121102A and 20190520B and their PRSs; the same is the case for FRB~20240114A, though with $P_i \approx 10$\,ms. However, \citet{rahaman2025} argue that the internal fields required by the model ($B_{\rm int} < 10^{13}$\,G) are too small to account for both the PRS and FRB emission, raising a viability issue for a purely rotationally powered origin.
    \item \textit{Magnetic decay powered.} \citet{rahaman2025} find that compact, luminous PRSs are more naturally sustained by magnetic-decay power with strong internal fields $B_{\mathrm{int}} \approx 10^{16}$--$10^{16.5}$\,G, slow nebular expansion $t_d\approx10$--$300$\,yr, and $P_i \gtrsim 10$\,ms. We note this $P_i$ requirement is somewhat in tension with the properties of FRB-PRS host galaxies, which favor the formation of millisecond magnetars with $P_i\lesssim 3\rm\,ms$ \citep{song2023}. Reconciling the \citet{rahaman2025} model with these expectations may require substantial early spin-down or a progenitor channel yielding intrinsically slower rotation.
\end{itemize}

An arising challenge for MWN models is the apparent stability of PRS emission over time, where gradual energy injection and nebular expansion should produce measurable flux evolution on decadal timescales. Long-term monitoring of FRB~20121102A-PRS shows no significant variability \citep{plavin2022, rhodes2023,bhardwaj2025}, though recent observations of FRB~20190520B-PRS present tentative evidence for flux decay and a low-frequency spectral break \citep{balaubramanian2025}.

Targeted broadband follow-up will help discriminate between, and assess the viability of, MWN models for FRB-PRS systems. In particular, MWNe are expected to exhibit broadband synchrotron spectra with a low-frequency self-absorption turnover ($\sim$$200$\,MHz), and potential high-frequency cooling breaks ($\sim$$ 150$--$200$\,GHz) that shift toward lower frequencies as the nebula expands \citep{rahaman2025}.

\subsubsection{Accretion-Powered Hypernebula}

In the hypernebula scenario, \citet{sridhar2022} propose that the PRS is powered by a compact nebula inflated by baryon-loaded outflows launched during a short-lived ($\lesssim 10^2-10^5$\,yr) episode of hyper-Eddington accretion onto a compact object, just before the compact object inspirals through the donor star's envelope. In this model, forward shocks or magnetic reconnection events (at $\sim$AU scales) due to flares launched along the jet funnel would power the ms-duration FRB \citep{Sridhar2021}. Particles energized at the large scale ($\lesssim$ pc) jet termination shock would emit synchrotron radiation largely in radio (PRS) and high-energy (TeV-PeV) neutrinos \citep{Sridhar+24}. Such powerful, turbulent outflows can naturally imprint the large host DM and variable RM observed for PRS-associated FRBs. Observations have also shown that metal-poor galaxies are more likely to host ultra-luminous X-ray sources (ULXs) \citep{zampieri2009,prestwich2013}, consistent with the presence of compact objects undergoing episodes of extreme accretion. We therefore investigate the viability of this model in light of our observations of \reighteen and \reighteen-PRS.

Following the formalism of \citet{sridhar2022}, we estimate the observable properties of the hypernebula for an assumed set of parameters; full details are provided in Appendix~\ref{ap:hypernebula}. Our calculations yield an active lifetime of the system $t_{\rm active} \sim  10^3\,{\rm yr}\,(M_\star/30\,M_\odot)(\dot{M}/10^5\,\dot{M}_{\rm Edd})^{-1}$, where $\dot{M}$ is the accretion rate, and $\dot{M}_{\rm Edd}\simeq1.4\times10^{19}\,{\rm g\,s^{-1}}$ is the Eddington accretion rate for an assumed $10\,M_\odot$ BH. Unlike MWN models, the radio emission is expected to plateau for a long time, or even rise/decrease slightly in the early/late stages of the nebula's evolution, depending on the system's parameters. This is consistent with the lack of significant flux evolution seen in the PRSs of FRBs~20121102A and 20190520B \citep{plavin2022,rhodes2023,bhardwaj2025,balaubramanian2025}. 

Powering the peak burst luminosities of \reighteen (derived using FRB fluxes from \citealt{fonseca2020}) requires an accretion rate of $\dot{M}=10^{5}\dot{M}_{\rm Edd}$. The outflowing slow winds (velocities $v_{\rm w}\sim0.01c$) from such an accretion disk would drive a forward shock into the CSM, inflating an expanding shell. The material in the expanding shell of the hypernebula contributes DM$_{\rm host}\approx1100\,{\rm pc\,cm^{-3}}$ at $t_{\rm DM}=7$\,yr. At this age, the size of the expanded shell is $R_{\rm sh}=v_{\rm w}t_{\rm DM} \simeq 0.02\,{\rm pc}$, with a smaller radio-emitting nebula confined within it; this is consistent with the upper limit on the PRS's transverse physical size of $23.1$\,pc. At the same age, the model also reproduces the observed PRS radio luminosity of $L_{1.4\rm\,GHz}\approx7.4\times10^{28}\,{\rm erg\,s^{-1}\,Hz^{-1}}$ (when accounting for cooling losses) for a jet speed of $v_{\rm j}=0.12\,c$. The maximum $| \rm RM|$ expected from the plasma is ${\rm |RM|_{max}\sim10^7\,rad\,m^{-2}}$, which comfortably encompasses our observed values of $\sim$ $4,000$--$5,000\,{\rm rad\,m^{-2}}$. In combination, these results demonstrate that the hypernebula framework can reproduce the observed properties of \reighteen and its PRS under plausible assumptions. We note that this is just one of the parameter combinations that can reproduce the observables, and is not a unique fit.

A challenge for the model is the brevity of its active phase ($\lesssim 10^2-10^5$\,yr), which must be reconciled with the observed $\sim$$5\%$ incidence of PRSs among repeating FRBs. The short lifetime implies that either the formation rate of such binaries is relatively high, or that FRB activity is preferentially triggered during the hypernebula phase. Future observations will be crucial for testing this framework, and better constraining the system's parameters. Sustained flux monitoring will reveal whether \reighteen-PRS plateaus, brightens, or decays, while high-frequency VLBI observations will further constrain its transverse physical size. The detection of a persistent X-ray counterpart or extended radio lobes would provide strong evidence for an accretion-powered hypernebula, though these signatures would only be detectable in exceptionally nearby systems.

\subsubsection{Wandering MBH}

The ``wandering'' (i.e., off-nuclear) MBH model posits that the PRS is powered by low-Eddington accretion onto a MBH ($10^4$--$10^7\,M_\odot$), while the FRB arises from a separate compact object --- such as a magnetar --- embedded in the BH's circumnuclear ($\sim$$0.1$--$10$\,pc) magneto-ionic environment. Low-Eddington accretion onto a MBH could generate a PRS via a small-scale jet or outflow, without demanding energetic youth \citep{annathomas2023,dong2024dwarf}. These sources would appear as compact, quasi-steady radio emitters offset from the host-galaxy center --- though in dwarf galaxies, the center is less well-defined.

Several observational signatures are consistent with this scenario. The large and evolving RMs and extragalactic DMs of PRS-associated FRBs are similar to those of the Galactic-center magnetar, PSR~J1745$-$2900, which is embedded in the magneto-ionic environment of an accreting (super-)MBH \citep{desvignes2018}. The low-level variability seen in FRB~20121102A-PRS over decadal timescales \citep{plavin2022, rhodes2023, bhardwaj2025} is also more characteristic of a steady AGN jet than of a rapidly evolving MWN.

Targeted radio surveys have uncovered analogues of FRB PRSs in nearby dwarf galaxies. \citet{reines2020} reported 20 off-nuclear luminous radio sources in nearby ($z < 0.055$) dwarf galaxies that are too bright to be supernova remnants or \ion{H}{2} regions. They proposed that these sources are accreting MBHs, wandering outside the nucleus. Follow-up studies by \citet{eftekhari2020} and \citet{dong2024dwarf} found that such sources closely match FRB PRSs in radio luminosity, spectrum and host environment. If the PRS is powered by a MBH, its radio luminosity would place it on the fundamental plane of accreting BHs, consistent with known low-luminosity AGNs \citep{mezcua2018b}.

A challenge for this model is its reliance on a chance coincidence between the FRB-emitting compact object and the accreting MBH. All confirmed PRS-associated FRBs exhibit above-average burst activity, and it is unclear why active sources would preferentially reside near a MBH. Moreover, deep X-ray limits for \ronetwin's PRS place stringent constraints on its accretion-powered emission, implying that if it is powered by a wandering MBH, the accretion must be highly radiatively inefficient \citep{sydnor2025}. At present, the wandering MBH interpretation rests largely on circumstantial evidence, bolstered by the observed long-term stability of PRS emission. High-sensitivity X-ray observations could directly probe accretion and, when combined with PRS radio luminosities, test consistency with the fundamental plane of accreting BHs. Additionally, VLBI imaging of nearby PRSs could further discriminate between a BH jet and a MWN. Finally, RM measurements of PRSs offer a complementary diagnostic: in the wandering MBH model, the $|\rm RMs|$ of the PRS and FRBs should be unrelated, reflecting their independent origins.

\subsubsection{Other Models}

Beyond the models discussed above, several alternative mechanisms have been proposed to explain the nature of PRSs. These include PWNe, off-axis afterglows from SLSNe/LGRBs, and long-lived radio emission from the interaction of supernova ejecta with the CSM.

A PWN origin is less favorable on energetic grounds. PWNe typically have radio luminosities of order $\sim$$ 10^{34}\rm\,erg\,s^{-1}$ \citep{gaensler2006}, several orders of magnitude below the $\sim$$10^{39}\rm\,erg\,s^{-1}$ luminosities of FRB PRSs. Off-axis SLSNe/LGRB afterglows and supernova-CSM interaction models remain viable, but they face challenges. A detailed interpretation of \reighteen in the context of these models is presented in Section 6 of \citet{ibik2024prs}.

\section{Conclusion}

We have presented the milliarcsecond-precision localization of \reighteen and confirmed its association with the PRS candidate identified by \citet{ibik2024prs}. Our key findings are as follows:

\begin{itemize}
    \item Combining five bursts, we localize \reighteen to the position:\\
    $\indent {\rm \alpha_{\rm FRB}\,(J2000)} = 19^{\rm h}39^{\rm m}05\fs8919 \pm4.9\rm\,mas,$\\
    $\indent {\rm \delta_{\rm FRB}\,(J2000)} = +59\degr19\arcmin36\farcs828 \pm5.2\rm\,mas.$\\
    referenced to the ICRF.
    \item We detect a compact, luminous radio source in the EVN continuum data of \reighteen, confirming the PRS proposed by \citet{ibik2024prs}. We verify its spatial coincidence with \reighteen and constrain their projected transverse physical offset to $< 26$\,pc. Visibility-domain model-fitting constrains the angular size of the PRS to $< 9.8\rm\,mas$, corresponding to a transverse physical size of $< 23\rm\,pc$.
    \item We demonstrate that \reighteen exhibits a large, time-variable RM$_{\rm rest} = 5038~\pm~14$ -- $6,441~\pm~31\rm\,rad\,m^{-2}$, while \reighteen-PRS has a luminosity of $L_{\rm 1.4~GHz} = (7.4\pm1.5)~ \times ~ 10^{28}\rm\,erg\,s^{-1}\,Hz^{-1}$. The system occupies an intermediate position on the empirical $L_{\nu}-\rm~|RM|$ relation \citep{yang2020}, between hyper-active, ultra-magnetized FRB-PRS systems (\rone, \ronetwin) and the fainter, lower-RM candidate \ronefourseven.
    \item Deep photometry and spectroscopy confirm that \reighteen resides in a low-metallicity ($12+\log(\rm O/H) = 7.95 \pm 0.01)$ dwarf galaxy with a very high sSFR $\sim$$2.3 \times 10^{-8}\rm\,yr^{-1}$, similar to the hosts of other PRS-associated FRBs. The concentration of FRB-PRS systems in the same chemically primitive, gas-rich galaxies that give rise to SLSNe and LGRBs strongly implies a metal-poor, rapidly spinning massive progenitor source.
    \item Our DM decomposition analysis yields a $90\%$-confidence lower limit $\rm DM_{host,rest} > 1228.7\,pc\,cm^{-3}$ and median DM$_{\rm host,rest} = 1275.0\rm\,pc\,cm^{-3}$, the highest known host contribution among FRBs so far. We find that FRB-PRS systems, in general, exhibit higher-than-average DM$_{\rm host}$ values and posit that this may be related to FRB-PRS systems being embedded in star-forming knots within their gas-rich hosts.
    \item FRB-PRS systems are emerging as a subclass of FRBs with distinct characteristics. We discuss these commonalities in the context of three scenarios: (i) evolutionary phase, wherein all repeating FRBs pass through a bright and short-lived PRS phase shorty after birth; (ii) environmental selection effect, wherein only FRBs born in the most metal-poor, gas-rich dwarf galaxies develop luminous PRSs; or (iii) distinct engine, wherein FRB-PRS systems might be powered by a source that is not universal to the broader FRB population. Though the evolutionary-phase interpretation seems less likely, deciding among these possibilities remains an open question and will require the discovery and continued monitoring of more FRB-PRS systems.
\end{itemize}

With the addition of \reighteen, the growing subclass of PRS-associated FRBs offers a unique window into the most extreme magneto-ionic environments of repeating FRBs. The leading PRS models --- MWNe, hypernebulae, and wandering MBHs --- are each well-motivated by current observations and consistent with the host environments in which FRB-PRS systems are predominantly found. Discriminating among these models will require targeted, long-term observations. In particular, we highlight the importance of: (i) broadband spectral monitoring to identify turnover frequencies or cooling breaks diagnostic of nebular evolution; (ii) continued VLBI imaging of nearby PRSs to resolve structure on $\lesssim 0.1$\,pc scales; (iii) sustained DM, RM and PRS flux monitoring to track temporal evolution; (iv) polarization measurements of PRS emission to determine whether the FRB and PRS share a common engine; and (v) the discovery and long-term study of new FRB-PRS systems to build a statistical sample. Such efforts will be crucial for establishing the place FRB-PRS systems hold within the broader FRB population.

\begin{acknowledgements}
The European VLBI Network is a joint facility of independent European, African, Asian, and North American radio astronomy institutes. Scientific results from data presented in this publication are derived from the following EVN project codes: EK050 and EK051.
We thank the directors and
staff of the participating telescopes for allowing us to observe with
their facilities.
This work is based in part on observations carried out using the 32 m radio telescope operated by the Institute of Astronomy of the Nicolaus Copernicus University in \torun (Poland) and supported by a Polish Ministry of Science and Higher Education SpUB grant.
S.B. is supported by a Dutch Research Council (NWO) Veni Fellowship (VI.Veni.212.058).
M.R.D. acknowledges support from the NSERC through grant No. RGPIN-2019-06186, the Canada Research Chairs Program, the Canadian Institute for Advanced Research (CIFAR), and the Dunlap Institute at the University of Toronto.
The AstroFlash research group at McGill University, University of Amsterdam, ASTRON, and JIVE acknowledge support from: a Canada Excellence Research Chair in Transient Astrophysics (CERC-2022-00009); an Advanced Grant from the European Research Council (ERC) under the European Union’s Horizon 2020 research and innovation programme (`EuroFlash'; Grant agreement No. 101098079); and an NWO-Vici grant (`AstroFlash'; VI.C.192.045).
B.M. acknowledges financial support from the State Agency for Research of the Spanish Ministry of Science and Innovation, and FEDER, UE, under grant PID2022-136828NB-C41/MICIU/AEI/10.13039/501100011033, and through the Unit of Excellence Mar\'ia de Maeztu 2020--2023 award to the Institute of Cosmos Sciences (CEX2019-000918-M).
Z.P. is supported by an NWO Veni fellowship (VI.Veni.222.295).
N.S. acknowledges support from the Simons Foundation (grant MP-SCMPS-00001470).
F.K. acknowledges support from Onsala Space Observatory for  the  provisioning of its facilities/observational support. The Onsala Space Observatory national research infrastructure is funded through Swedish Research Council grant No 2017-00648.
C.L. acknowledges support from the Miller Institute for Basic Research at UC Berkeley.
K.W.M. holds the Adam J. Burgasser Chair in Astrophysics and is supported by NSF grant 2018490.
D.M. acknowledges support from the French government under the France 2030 investment plan, as part of the Initiative d'Excellence d'Aix-Marseille Universit\'e -- A*MIDEX (AMX-23-CEI-088).
K.N. is an MIT Kavli Fellow.
A.P. is funded by the NSERC Canada Graduate Scholarships -- Doctoral program.
A.B.P. is a Banting Fellow, a McGill Space Institute~(MSI) Fellow, and a Fonds de Recherche du Quebec -- Nature et Technologies~(FRQNT) postdoctoral fellow.
P.S. acknowledges the support of an NSERC Discovery Grant (RGPIN-2024-06266).
\\

\facilities{EVN}\\

\software{FX Software Correlator \citep{kempeima2015}, CASA \citep{mcmullin2007casa,casa2022,casa2022fringe}, Difmap \citep{difmap1997}, Astropy \citep{astropy}, Sigproc \citep{lorimer2011}.}\\

\textit{Data availability:} Uncalibrated visibilities of \reighteen (both the bursts and the associated PRS) and its calibration sources can be downloaded from the JIVE/EVN archive, \url{https://archive.jive.eu}, under project codes EK050C, EK050D, EK050F, EK050G, EK051A, EK051B, and EK051D. Filterbank files, calibrated burst visibilities, dirty maps fits files, and the scripts that made Figures~\ref{fig:PRS}, \ref{fig:PRSFRB}, \ref{fig:DMz} and \ref{fig:rm-evolution} can be accessed in our Zenodo reproduction package: 10.5281/zenodo.17582142. Due to the large file sizes, the burst baseband data (i.e., raw voltages) and calibrated continuum visibilities will be made available by the authors upon reasonable request.

\end{acknowledgements}

\appendix

\section{Burst Dynamic Spectra}\label{sec:familyplot}

\begin{figure*}[!ht]
\plotone{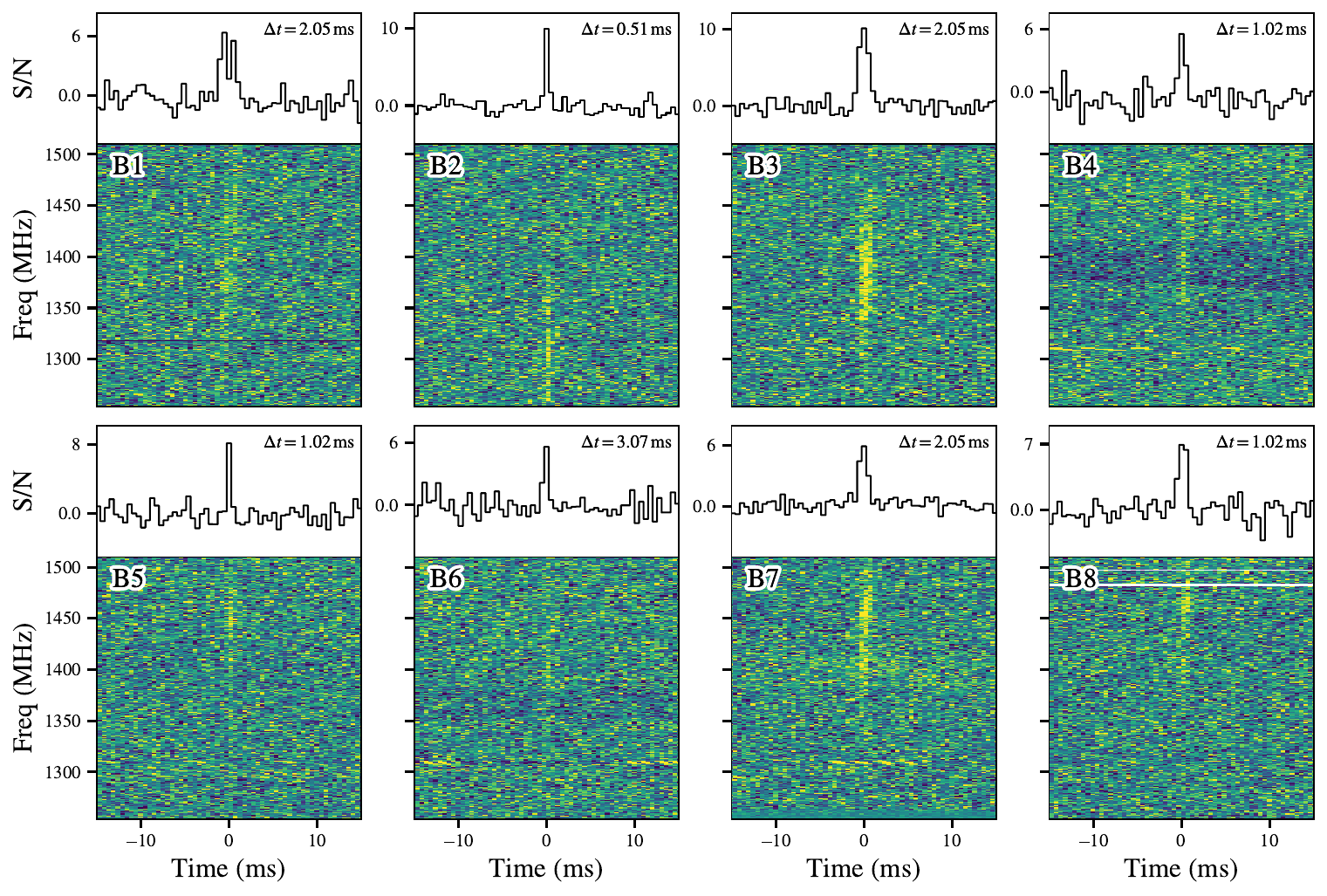}
\label{fig:familyplot}
\caption{Temporal profiles (top) and dynamic spectra (bottom) of the eight \reighteen bursts detected with the Ef radio telescope. Each burst is coherently dedispersed to a DM of $1379.0\rm\,pc\,cm^{-3}$ and is shown with a frequency resolution of $1$\,MHz; the time resolutions are shown in the top right corners of each panel. Horizontal white bands correspond to frequency channels that have been flagged due to RFI. The limits of the color map have been set to the $20^{\mathrm{th}}$ and $98^{\mathrm{th}}$ percentile of each dynamic spectrum. We note some of the wider bursts appear slightly over-dedispersed; future work will analyze all bursts at a higher time resolution to place tighter constraints on their DM.}
\end{figure*}

Figure~\ref{fig:familyplot} shows all temporal profiles and dynamic spectra of the bursts detected by the Ef telescope in this campaign.

\section{Polarimetry}\label{sec:polarimetry}
We perform polarimetric calibration of the burst data from Ef (circular basis) using a test pulsar instead of a noise diode \citep[similar to, e.g.,][]{kirsten2021}. During each of our EVN observations, one/two scans of a few minutes were conducted on the pulsar~B2255+58. For each pulsar scan, we measure the instrumental delay between the polarization hands by first derotating the Q and U data using the known RM of the pulsar ($-323.5$\,rad\,m$^{-2}$), and then search for the delay (in the range of $-20$ to $+20$\,ns) which maximizes the linear polarized flux. For each trial the linear polarization vector ($Q + iU$) of the pulsar data is thus multiplied with a phase correction: 
\begin{equation}
   \phi_{\rm corr} = \text{exp}( -2i(c^2\text{RM}/\nu^2 + \pi\nu\tau ))
\end{equation}

\noindent where RM = $-323.5$\,rad\,m$^{-2}$ \citep{ATNF}\footnote{\url{https://www.atnf.csiro.au/research/pulsar/psrcat/}}, $c$ is the speed of light, $\nu$ is the observing frequency and $\tau$ is the cable/instrumental delay between the polarization hands. The delays measured using this method range from $\sim$$1-4\,$ns and vary by a few percent between pulsar observations conducted on the same day ($\sim$$10$\,hr apart). In order to replicate the polarimetric profile of PSR~B2255+58 in the EPN database \citep{Gould_1998_MNRAS}, we assume a flip between the polarization hands (equivalent to changing convention).

We then derotate the linear popularization vector of each burst from \reighteen by multiplying with a phase correction using the measured delay from the pulsar. Thereafter, we search a range of RM values between $-10,000$ and $+10,000\rm\,rad\,m^{-2}$ to determine at which RM trial the linear polarized flux peaks. This is done through RM synthesis, which produces the Faraday dispersion function (FDF) --- a representation of how polarized emission is distributed across Faraday depth \citep{brentjens2005}. We report RM$_{\text{FDF}}$ as the depth corresponding to the strongest peak in this distribution, with values tabulated in Table~\ref{tab:burstproperties}. In our analysis, we assume that the sources are Faraday simple (i.e., described by a single Faraday component). We note that this assumption may lead to a slight underestimation of the RM uncertainties in the presence of Faraday complexity. However, all FDF detections in our burst sample exceed S/N = 6, with B5 being the faintest at S/N = 5.8.

Three of the bursts (B2, B3 and B8) were sufficiently bright to see the cyclical intensity fluctuations induced by Faraday rotation in Q and U across the observing band. For these bursts, we also perform a QU-fit \citep{purcell2020} to Stokes Q/L and U/L (where $L = \sqrt{Q^2 + U^2}$ is the total linear polarization) over the spectral extent of the burst. We use the following equations: 

\begin{equation}
    Q/L = \text{cos}(2(c^2\mathrm{RM}/\nu^2 + \nu\pi\tau + \phi))
\end{equation}
\begin{equation}
    U/L = \text{sin}(2(c^2\mathrm{RM}/\nu^2 + \nu\pi\tau + \phi))
\end{equation}
\\
\noindent with $\phi$ referring to an additional phase term related to the polarization position angle at infinite frequency. The delay is fixed to the delay which maximized the linear polarized flux of the pulsar observation. These RM measurements (RM$_{\text{QU}}$ in Table~\ref{tab:burstproperties}) are consistent with those obtained by maximizing linear polarization. 

\begin{figure*}
    \centering
\includegraphics[width=.49\textwidth]{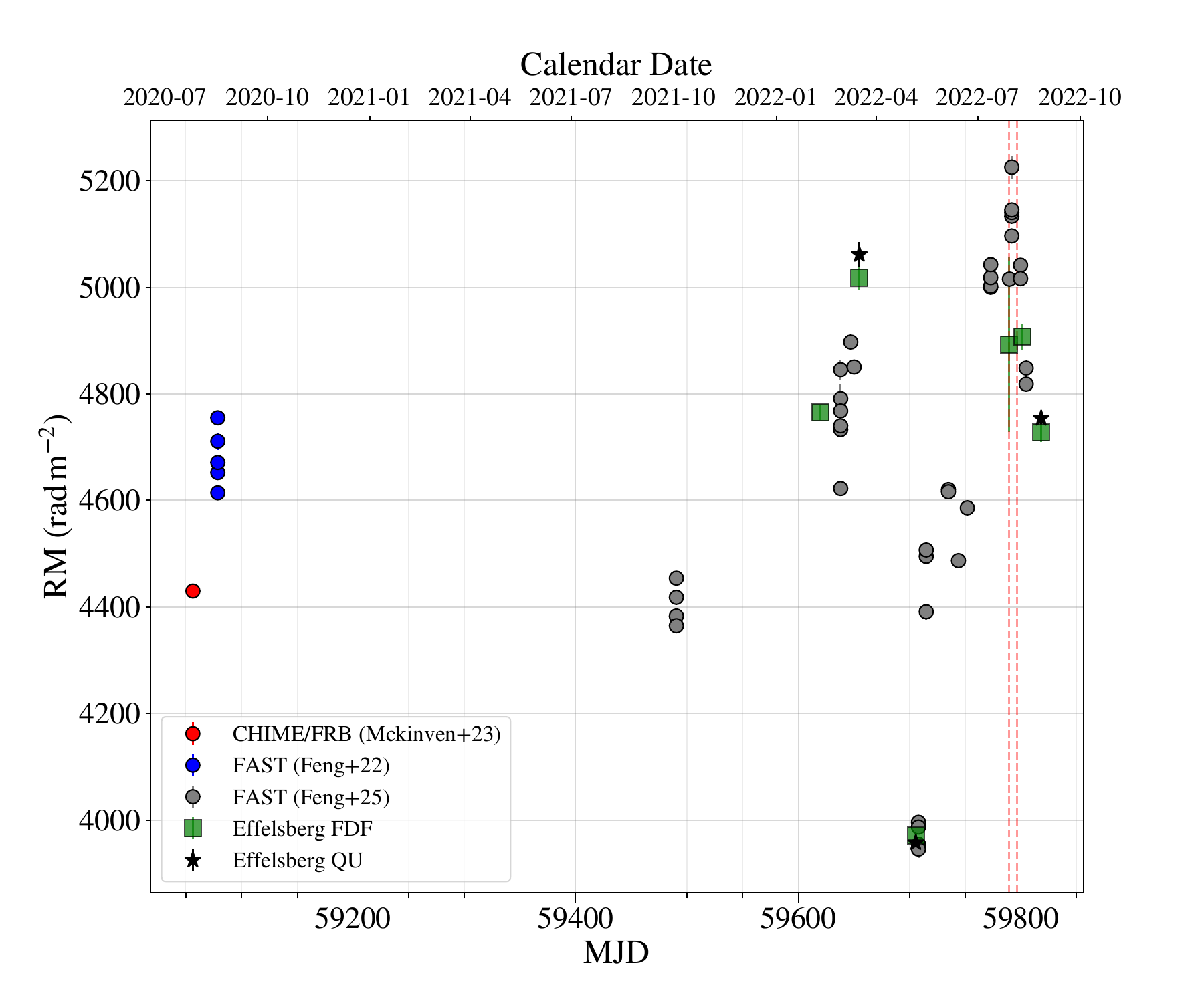}
\includegraphics[width=.49\textwidth]{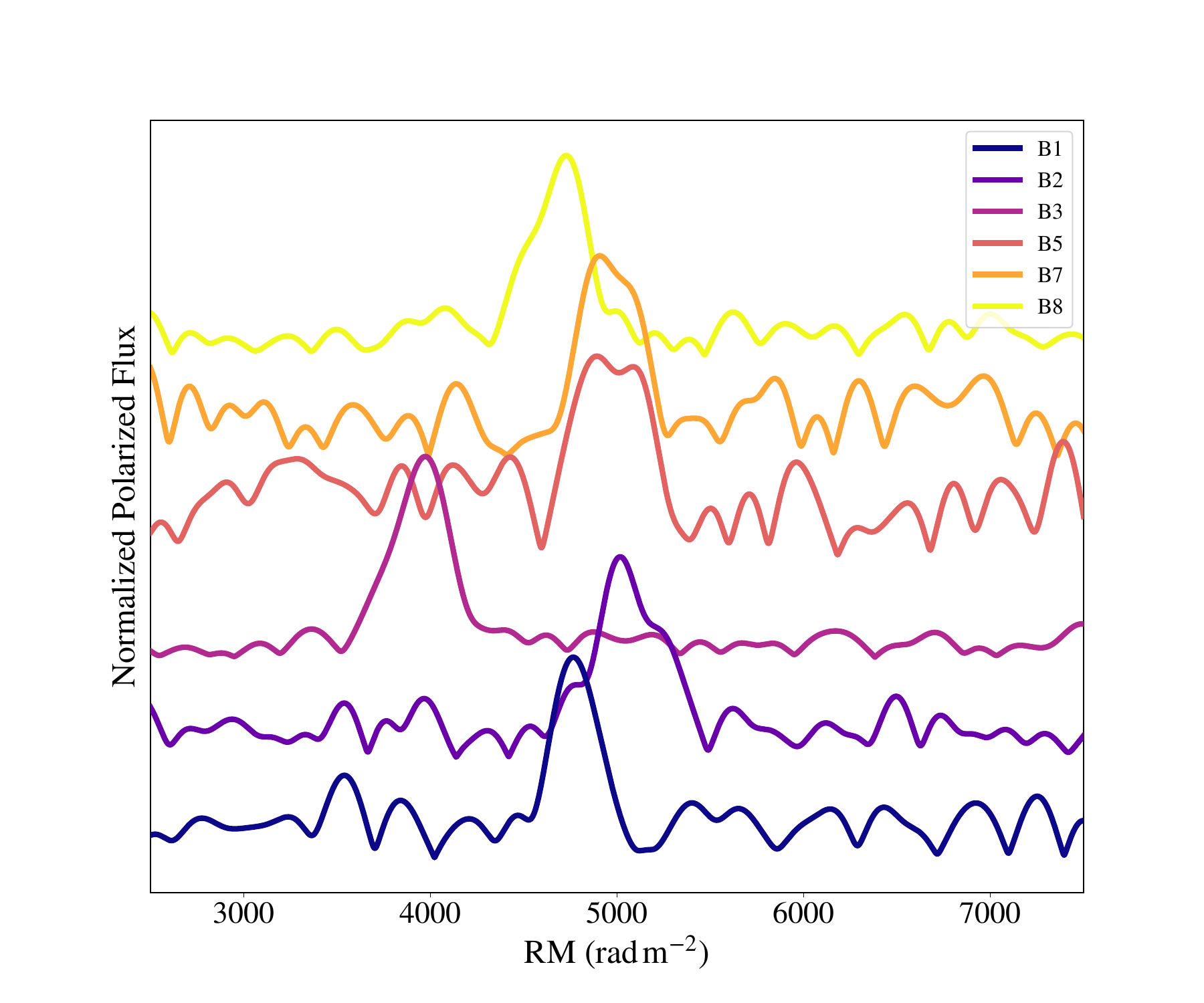}
\caption{Left: The RM evolution of \reighteen. Previous RM measurements taken from (\citealt{feng2022rm}; blue circles), (\citealt{mckinven2023rm}; red circle) and (\citealt{feng2025}; grey circles). The green squares indicate RM measurements obtained using a brute-force search to maximize linear polarization. The error bars are calculated as FWHM/(S/N) of the brightest peak in the FDF. The black stars indicate the RM measurements obtained from a QU-fit for the bursts that were sufficiently bright. Vertical dashed lines indicate the MJDs of B4 and B6 that were too low S/N to obtain an RM measurement. Right: Faraday dispersion functions of the \reighteen bursts detected with Ef. The instrumental delay has already been accounted for, so the shift in the peak indicates significant variation in the RM of the source.}
\label{fig:rm-evolution}
\end{figure*}

\section{DM Decomposition Method}\label{sec:DMappendix}

Here we adopt a deliberately conservative approach designed to avoid overestimating the ${\rm DM_{host}}$ contribution of \reighteen. The rest frame host contribution of an FRB can be obtained by rearranging Equation~\ref{eq:DM} as follows:

\begin{equation}
\mathrm{DM_{host,rest}} = (1+z) \left( \mathrm{DM_{obs}} - \mathrm{DM_{MW}} - \big< \mathrm{DM_{cosmic}} (z)\big> - \sum_i~\frac{\mathrm{DM}_{\mathrm{halo},i}}{1+z_i} \right),
\end{equation}

where $\rm DM_{MW}$ is the Galactic contribution (including the disk and halo), $\big< \mathrm{DM_{cosmic}}\big>$ is the mean IGM component and $\mathrm{DM}_{\mathrm{halo},i}$ denotes the contribution from intervening halos along the LoS.

The MW disk contribution is estimated using the NE2001 electron density model \citep{ne2001,ne2001p}. For the FRB's Galactic coordinates $(l, b) = (92.21\degr, +19.50\degr)$, the model returns $\rm DM_{MW,disk} = 78.0\,pc\,cm^{-3}$. We adopt a uniform prior centered on this value $\rm DM_{MW,disk} \sim \mathcal{U}(78.0 \pm 0.2\,DM_{\rm MW,disk})\,pc\,cm^{-3}$, corresponding to a $\pm20$\% uncertainty on the NE2001 estimate. The MW halo component is less certain; based on constraints on the circum-galactic medium \citep{cook2023dm, ravi2025} we assume a conservative $\rm DM_{MW,halo} \sim \mathcal{U}(10, 111)\,pc\,cm^{-3}$. Combining these components, we obtain a total MW contribution of $\rm DM_{MW} = DM_{MW,disk} + DM_{MW,halo}$.

Following the formalism of \citet{macquart2020}, the mean IGM contribution is:

\begin{equation}
\big< \mathrm{DM_{cosmic}} (z)\big> = \frac{3 c H_0 \Omega_b f_{\rm IGM}}{8 \pi G m_p} \int_0^z \frac{(1 + z')f_e(z')}{E(z')} \, dz',
\end{equation}

with $H_0 = 67.4\rm\,km~s^{-1}~Mpc^{-1}$, $\Omega_b = 0.049$,  $f_{\rm IGM} = 0.85$ and $f_e(z') \approx 7/8$ is the free electron number per baryon in the Universe \citep{planck18}. At the redshift of the host, $z = 0.12817$, we find $\big< \mathrm{DM_{cosmic}} \big> \approx 110\rm\,pc\,cm^{-3}$. By explicitly including a $\rm DM_{halos}$ term, we effectively account for fluctuations from intervening over-densities in the IGM, and thus assume that the residual scatter in $\big< \mathrm{DM_{cosmic}} (z)\big>$ is negligible compared to the variance introduced by halos.

FRB~20190520B has demonstrated that exceptional LoS scenarios, such as an FRB passing through multiple galaxy clusters, can inflate its DM \citep{lee2023flimflam}. In the case of \reighteen, we cross-match the burst position with the optically selected cluster catalogue of \citet{wen2024}. We rule out \reighteen intersecting a cluster or group of mass $\gtrsim 5\times10^{13}\rm~M_{\odot}$ within $2.8\times r_{500}$, where $r_{500}$ corresponds to the radius, $r$, within which the mean density is 500 times the critical density \citep{wen2024}.

To account for additional intervening halos, we draw from the methods outlined in \citet{prochaska2019,mcclintock2019} and \citet{simha2023flimflam}. Since we ruled out intersecting halos $\gtrsim 5\times10^{13}\rm\,M_{\odot}$, we adopt three mass bins for our DM$_{\rm halo}$ analysis  -- dwarf galaxies ($\rm  10^{8-11}\,M_{\odot}$), L$^*$ galaxies ($\rm  10^{11-13}\,M_{\odot}$) and galaxy groups ($\rm 1 - 5 \times 10^{13}\,M_{\odot}$). Each halo is assigned a hot-gas fraction, $f_{\rm hot}$; $0.02f_b$ for dwarfs, $0.15f_b$ for L$^*$ galaxies and $0.3f_b$ for groups where $f_b = \Omega_b/\Omega_m \approx 0.158$ is the cosmological baryon fraction \citep{planck18}.

Halo concentrations, $c$, are calculated with the \texttt{Colossus} concentration module \citep{colossus} using the \citet{diemer19} mass-concentration relation, while virial radii $r_{200}$ are obtained by inverting the spherical-overdensity definition:

\begin{equation}
    r_{200} = \left( \frac{3M_{200}}{4\pi\, 200 \rho_{\rm crit} \left(z \right) } \right)^{1/3},
\end{equation}

where $\rho_{\rm crit}$ is the critical density ($\rm kg~m^{-3}$). Assuming the hot gas component traces a dark-matter Navarro-Frenk-White (NFW) profile \citep{nfw}, the electron density is:

\begin{equation}
    n_e(r) = \frac{n_0}{x(1+x)^2},
\end{equation}

with $x=r/r_s$ where $r_s = r_{200}/c$ is the scale radius of the NFW profile. The central normalization, $n_0$, is solved analytically so that the integral of $n_e$ out to $r_{200}$ equals the allocated hot-gas mass. For a sight-line intersecting a halo at impact parameter $b < r_{200}$, our analysis carries out a 1-D line integral along the chord length $l_{\rm max} = \sqrt{r^2_{200} -b^2}$. The resulting column density is converted to DM units and divided by $(1+z)$ to place it in the observer frame for consistency with the rest of the analysis.

We populate the LoS with intervening halos using the \citet{tinker08} halo mass function as implemented in \texttt{Colossus}. Masses are drawn via inverse-transform sampling of the pre-tabulated cumulative function; $z$ is sampled uniformly in comoving distance to preserve the expected comoving number density, $n_c(M,z)$. Using the formalism of \citet{prochaska2019}, we compute the expected number of intercepts for each mass bin as:

\begin{equation}
\big< N \big> = \int_0^z n_c(M,z)\sigma(M) \frac{dX}{dz} \, dz,
\end{equation}

where $\sigma (M) = \pi R^2$ is the cross-sectional area of the halo (Mpc$^2$), and $\frac{dX}{dz}=(1+z)^2/E(z)$ with $E(z)$ referring to the dimensionless Hubble parameter \citep{hogg1999}. For simplicity, we assume a representative halo mass and radii per mass bin --- $M_{200} \sim 10^{9}\,M_{\odot}$, $R_{200} = 60\rm\,kpc$ for dwarf galaxies; $M_{200} \sim 10^{12}\,M_{\odot}$, $R_{200} = 250\rm\,kpc$ for L$^*$ galaxies; $M_{200} \sim 3\times 10^{13}\,M_{\odot}$, $R_{200} = 500\rm\,kpc$ for groups \citep{giodini2009, kravtsov2013, callingham2019}. This results in $\big< N_{\mathrm{dwarf}} \big> = 4.14$, $\big< N_{\mathrm{L_*}} \big> = 0.13$, and $\big< N_{\mathrm{group}} \big> = 0.02$ at $z = 0.12817$. Assuming the number of intervening halos per mass bin follows a Poissonian distribution with $\lambda = \big<N\big>$, we simulate the expected DM per mass bin for $1,000,000$ Monte Carlo sight-lines. We then sum the contributions, returning both the total DM$_{\rm halos}$ and its breakdown by mass bin.

Finally, we combine the distributions of $\rm DM_{MW}$, $\big< \mathrm{DM_{cosmic}}\big>$, and $\rm DM_{halos}$ to compute the posterior on $\rm DM_{host}$. Note that, by adding an explicit $\mathrm{DM}_{\mathrm{halos}}$ term on top of $\big< \mathrm{DM_{cosmic}} \big>$, we are partially double-counting the intervening halo contribution. We retain this term to be conservative; our 90\%-confidence lower limit on $\rm DM_{host}$ is insensitive to the tail of the distribution, and adopting tighter constraints would only raise the inferred lower limit.

Our analysis finds a median $\rm DM_{host, obs} = 1130.0\,pc\,cm^{-3}$ and places a $90\%$-confidence lower limit on $\rm DM_{host,obs} > 1088.8\,pc\,cm^{-3}$ in the observer's frame; the largest known host contribution among repeaters. In the source rest frame, this corresponds to a median of $\rm DM_{host, rest} = 1275.0\,pc\,cm^{-3}$ and a $90\%$-confidence lower limit on $\rm DM_{host,rest} > 1228.7\,pc\,cm^{-3}$, indicating that \reighteen resides in an exceptionally dense environment.

\section{Additional Host Galaxy Information}\label{sec:HostAppendix}

\subsection{Emission Line Fluxes}\label{ap:emission}

In Table~\ref{tab:linefluxes} we list the emission line fluxes found for the host galaxy based on the analysis described in Section~\ref{sec:spectrum}. However, we note that the H$\beta$ and [\ion{O}{3}] line fluxes should be treated with caution. As noted in \cite{ibik2024prs}, the spectrum shows a very low (possibly unphysical) Balmer decrement of H$_{\alpha}$/H$_{\beta}$ $=$ 0.94. This is in stark contrast to the typical value of 2.87 for Case B recombination at $10,000$\,K \citep{Osterbrock2006} or even larger values expected when additional reddening is present. 
While it is possible to achieve Balmer decrements near unity with the emitting material has very high densities \citep[n$_e \gtrsim 10^{13}$\,cm$^{-3}$;][]{Drake1980,Levesque2014}, this could also indicate an issue with the original flux calibration of the spectrum. 

In particular, at the redshift of the host galaxy, both H$\beta$ and the [\ion{O}{3}] lines are close to the edge of where the sensitivity of the GMOS R400 grating falls off in the blue. Flux calibration uncertainties can therefore increase in this region (we note that the uncertainties quoted in Table~\ref{tab:linefluxes} are purely statistical). While photometry of the host galaxy can be used to correct the flux calibration (including a possible warping with wavelength), the H$\beta$ and the [\ion{O}{3}] lines lie just blue-wards of the observed r-band and we do not have compute spectral coverage for the g-band. We therefore opted to perform a single vertical offset for the final spectrum based on the observed i-band magnitude for the galaxy, as described in Section~\ref{sec:spectrum}. It is for this reason that we opt to use the PP04~N2 diagnostic (which relies only on emission lines located with in the observed i-band) as opposed to other diagnostics (such as the recent O3N2 calibration of \citealt{Hirschauer2018}, which has been used in other recent FRB studies but relies on all of the H$_{\alpha}$, [\ion{H}{2}], H$_{\beta}$, and [\ion{O}{3}] lines).

Finally, we note that even if the H$\beta$ and the [\ion{O}{3}] line fluxes are overestimated, we do not expect our conclusion that the host overlaps with star-forming galaxies (as opposed to AGNe) in the BPT diagram to change. Unlike several metallicity diagnostics, the BPT diagram relies solely on the ratios of lines located at similar wavelengths (\ion{O}{3} $\lambda$5007/H$_{\beta}$ and [\ion{N}{2}] $\lambda$6583/H$_{\alpha}$), which helps to mitigate uncertainties with either flux calibration or reddening. In particular, because the ratio of [\ion{N}{2}] $\lambda$6583 to H$_{\alpha}$ is so large (log([\ion{N}{2}]/H$_{\alpha}$) $=$ $-1.66$), we would require log([\ion{O}{3}]/H$_{\beta}$) $\gtrsim$1 in order for the galaxy to fall in the AGN portion of the BPT diagram. This, in turn, would require that the relative flux calibration between $\lambda$5007 and $\lambda$4861 to be off by more than a factor of 6 (whereas resolving the Balmer decrement discrepancy would only require the the relative flux calibration between H$_{\alpha}$ and H$_{\beta}$ --- a much larger wavelength range --- was off by a factor of 2--3).

\begin{table*}[h]
\centering
\caption{Measured Emission Line Fluxes for the Host Galaxy.}\label{tab:linefluxes}
\begin{tabular}{lc}
\hline
Emission Line & Flux (erg\,s$^{-1}$\,cm$^{-2}$) \\
\hline
H$_{\alpha}$ $\lambda$6563     & $(8.15 \pm 0.02) \times 10^{-16}$\\
H$_{\beta}$ $\lambda$4861      & $(8.69 \pm 0.02) \times 10^{-16}$\\
$[$\ion{N}{2}$]$ $\lambda$6583 & $(1.78 \pm 0.01) \times 10^{-17}$\\
$[$\ion{S}{2}$]$ $\lambda$6716 & $(3.88 \pm 0.01) \times 10^{-17}$\\
$[$\ion{S}{2}$]$ $\lambda$6731 & $(3.77 \pm 0.01) \times 10^{-17}$\\
$[$\ion{O}{3}$]$ $\lambda$4959 & $(9.20 \pm 0.02) \times 10^{-16}$ \\
$[$\ion{O}{3}$]$ $\lambda$5007 & $(1.75 \pm 0.01) \times 10^{-15}$\\
\hline
\multicolumn{2}{l}{Note: H$_{\beta}$ and [\ion{O}{3}] lines fluxes may be overestimated} \\
\multicolumn{2}{l}{due to a flux calibration issue in the blue portion of the} \\
\multicolumn{2}{l}{spectrum (see text for details).}
\end{tabular}
\end{table*}

\section{\reighteen in the Context of the Hypernebula Model}\label{ap:hypernebula}

We investigate the viability of the accretion-powered hypernebula model in light of our \reighteen observations. Note that this is just one of the parameter combinations of the hypernebula system that can reproduce the observables, and is not a unique fit.

Following the equations presented in \cite{sridhar2022}, we estimate the observable properties of the hypernebula assuming the following physical parameters: the jet magnetization (ratio of magnetic to plasma enthalpy density) $\sigma_{\rm j}=0.1$, the disk-wind to jet luminosity ratio $\eta = 0.1$, the fraction of the shock power that goes into heating the electrons $\epsilon_{\rm e} = 0.5$, and the mass of the companion accretor star $M_\star = 30\,M_\odot$. This sets the active lifetime of the system to be $t_{\rm active} \sim  10^3\,{\rm yr}\,(M_\star/30\,M_\odot)(\dot{M}/10^5\,\dot{M}_{\rm Edd})^{-1}$, where $\dot{M}$ is the accretion rate, and $\dot{M}_{\rm Edd}\simeq1.4\times10^{19}\,{\rm g\,s^{-1}}$ is the Eddington accretion rate for an assumed BH of mass $M_\bullet=10\,M_\odot$.

Let us consider the peak radio burst luminosity from \reighteen to be $10^{41-42}\,{\rm erg\,s^{-1}}$, adopting
FRB fluxes from \citealt{fonseca2020} and the host redshift of
$z = 0.12817$. Powering this would require an accretion rate of $\dot{M}=10^{5}\dot{M}_{\rm Edd}$. The outflowing slower winds (with a speed $v_{\rm w}\sim0.01c$) from such an accretion disk would drive a forward shock into the CSM with an assumed density $n \approx 10\,{\rm cm^{-3}}$. They would freely expand until a time:

\begin{equation}
t_{\rm free} =  424\,{\rm yr}\,\left(\frac{L_{\rm w,41}}{n_1}\right)^{1/2}\left(\frac{v_{\rm w}}{0.01\,c}\right)^{-2/5},
\end{equation}

where $L_{\rm w}=\dot{M}v_{\rm w}^2/2 = 6\times10^{40}\,{\rm erg/s}$ is the power of the outflowing winds. Here, we adopt the shorthand notation, $Y_{x} \equiv Y/10^x$ for quantities in cgs units. The contribution of the expanding material to the DM (assuming the material remains ionized), during the free expansion phase, is given by:
\begin{align}\label{eq:DM_sh}
    & \text{DM}_{\rm sh} \simeq \frac{M_{\rm sh}}{4\pi R^{2}m_{\rm p}} \approx
    \begin{cases}
          18\,{\rm pc\,cm^{-3}}\left(\frac{\dot{M}}{10^5\dot{M}_{\rm Edd}}\right) \left(\frac{v_{\rm w}}{0.01\,c}\right)^{-2} \left(\frac{t}{424\,{\rm yr}}\right)^{-1} & (t < t_{\rm free}) \\ 
          10\,{\rm pc\,cm^{-3}}\left(\frac{L_{\rm w,41}}{n_{1}}\right)^{1/5}\left(\frac{t}{424\,{\rm yr}}\right)^{3/5} & (t > t_{\rm free}),
    \end{cases}
\end{align}
where we take the mass in the expanding shell to be $M_{\rm sh}\sim\dot{M}t$, and $m_{\rm p}=1.67\times10^{-24}$\,g is proton's mass. We see that at a time $t_{\rm DM}=7$\,yr, the material in the expanding shell contributes a DM of $\rm DM_{host}\approx1100\,{\rm pc\,cm^{-3}}$, as seen from \reighteen. The other solution of $t=10^6\,{\rm yr}$ (for $t>t_{\rm free}$) is beyond the lifetime of the system, so we do not consider it. At this age, the size of the expanded shell is $R_{\rm sh}=v_{\rm w}t_{\rm DM} \simeq 0.02\,{\rm pc}$, with a smaller radio-emitting nebula confined within it; this is consistent with the upper limit on the PRS's transverse size of 23.1 pc. This model, at $t=t_{\rm DM}$, consistently also reproduces the observed PRS spectral luminosity of $L_\nu=7.4\times10^{28}\,{\rm erg\,s^{-1}\,Hz^{-1}}$ (at 1.4\,GHz) --- taking into account the effects of various cooling losses --- for a faster jet speed of $v_{\rm j}=0.12\,c$ (cf. Eqs.~29-42 of \citealt{sridhar2022}). The absolute maximum rotation measure expected from this plasma is ${\rm |RM|_{max}\sim10^7\,rad\,m^{-2}}$, which is consistent with the observed value of $\sim$$4700\,{\rm rad\,m^{-2}}$ (cf. Eq.~50 of \citealt{sridhar2022}).

\end{document}